%% file: main.tex
\documentclass[acmtog,nonacm]{acmart}
\AtBeginDocument{%
  }

\copyrightyear{2026}
\acmYear{2026}
\setcopyright{cc}
\setcctype{by}
\acmConference[SIGGRAPH Conference Papers '26]{Special Interest Group on Computer Graphics and Interactive Techniques Conference Conference Papers}{July 19--23, 2026}{Los Angeles, CA, USA}
\acmBooktitle{Special Interest Group on Computer Graphics and Interactive Techniques Conference Conference Papers (SIGGRAPH Conference Papers '26), July 19--23, 2026, Los Angeles, CA, USA}
\acmDOI{10.1145/3799902.3811085}
\acmISBN{979-8-4007-2554-8/2026/07}

\input{preamble}

\newcommand{\Gaussian}{\ensuremath{G}\xspace}
\newcommand{\mean}{\ensuremath{\mu}\xspace}
\newcommand{\cov}{\ensuremath{\Sigma}\xspace}
\newcommand{\opacity}{\ensuremath{o}\xspace}
\newcommand{\NumberOfGaussians}{\ensuremath{N}\xspace}
\newcommand{\pixel}{\ensuremath{d}\xspace}
\newcommand{\CameraPose}{\ensuremath{C}\xspace}
\newcommand{\CurrentCam}{\ensuremath{c}\xspace}
\newcommand{\NumberOfCams}{\ensuremath{M}\xspace}
\newcommand{\Times}{\ensuremath{T}\xspace}
\newcommand{\CurrentTime}{\ensuremath{t}\xspace}
\newcommand{\image}{\ensuremath{I}\xspace}
\newcommand{\pixels}{\ensuremath{X}\xspace}
\newcommand{\NumberOfPixels}{\ensuremath{D}\xspace}
\newcommand{\CurrentPixel}{\ensuremath{d}\xspace}
\newcommand{\NumberOfFrames}{\ensuremath{K}\xspace}

\newcommand{\Ray}{\ensuremath{R}\xspace}
\newcommand{\CurrentRay}{\ensuremath{r}\xspace}

\newcommand{\PrimitiveColor}{\ensuremath{s}\xspace}

\newcommand{\components}{\ensuremath{\delta}\xspace}
\newcommand{\parameters}{\ensuremath{\theta}\xspace}

\newcommand{\score}{\ensuremath{score}\xspace}

\begin{document}

\title{Graphical X Splatting (GraphiXS): A Graphical Model for 4D Gaussian Splatting under Uncertainty}


\author{Do\u{g}a Y\i lmaz}
\orcid{0000-0002-2268-7136}
\affiliation{%
  \institution{University College London}
  \city{London}
  \country{United Kingdom}}
\email{doga.yilmaz@ucl.ac.uk}

\author{Jialin Zhu}
\orcid{0000-0002-1826-6566}
\affiliation{%
  \institution{Baidu Inc.}
  \city{Beijing}
  \country{China}}
\email{misaliet@outlook.com}

\author{Deshan Gong}
\orcid{0009-0002-2516-9542}
\affiliation{%
  \institution{The University of Hong Kong}
  \city{Hong Kong}
  \country{China}}
\email{deshan@hku.hk}

\author{He Wang}
\authornote{Corresponding author.}
\orcid{0000-0002-2281-5679}
\affiliation{%
  \institution{University College London}
  \city{London}
  \country{United Kingdom}}
\email{he_wang@ucl.ac.uk}

%
\renewcommand{\shortauthors}{Yilmaz et al.}

\begin{abstract}
We propose a new framework to systematically incorporate data uncertainty in Gaussian Splatting. Being the new paradigm of neural rendering, Gaussian Splatting has been investigated in many applications, with the main effort in extending its representation, improving its optimization process, and accelerating its speed. However, one orthogonal, much needed, but under-explored area is data uncertainty. In standard 4D Gaussian Splatting, data uncertainty can manifest as view sparsity, missing frames, camera asynchronization, \etc. So far, there has been little research to holistically incorporating various types of data uncertainty under a single framework. To this end, we propose Graphical X Splatting, or \textit{GraphiXS}, a new probabilistic framework that considers multiple types of data uncertainty, aiming for a fundamental augmentation of the current 4D Gaussian Splatting paradigm into a probabilistic setting. GraphiXS is general and can be instantiated with a range of primitives, \eg Gaussians, Student's-t. Furthermore, GraphiXS can be used to `upgrade' existing methods to accommodate data uncertainty. Through exhaustive evaluation and comparison, we demonstrate that GraphiXS can systematically model various uncertainties in data, outperform existing methods in many settings where data are missing or polluted in space and time, and therefore is a major generalization of the current 4D Gaussian Splatting research.
\end{abstract}

\ccsdesc[500]{Computing methodologies~Machine learning approaches}
\ccsdesc[500]{Computing methodologies~Rendering}


\begin{teaserfigure}
  \includegraphics[width=\textwidth]{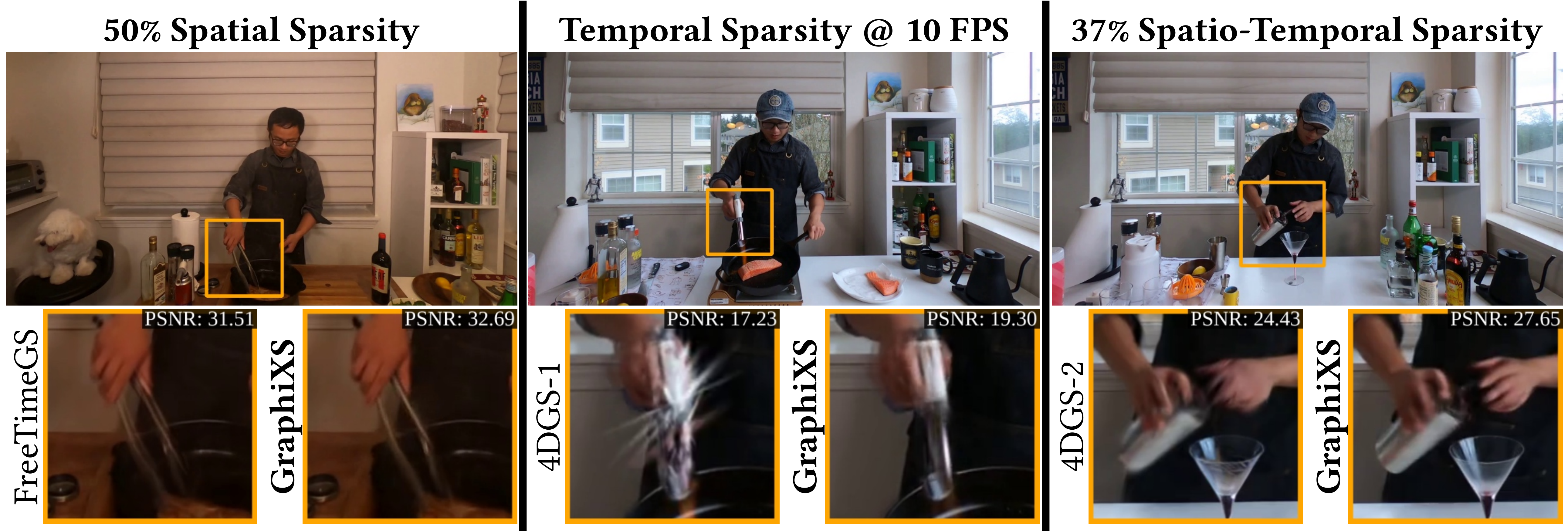}
  \caption{GraphiXS outperforms existing 4DGS methods under various types of data uncertainty. Left: 50\% cameras missing; Middle: 10 FPS low-speed cameras; Right: 37\% random frames missing.}
  \label{fig:teaser}
\end{teaserfigure}


\maketitle

\begin{bibunit}[ACM-Reference-Format]

\section{Introduction}
\label{sec:intro}

As the latest paradigm of 3D reconstruction and neural rendering, Gaussian Splatting (GS) has served as a fundamental component of many systems~\cite{zhou2024drivinggaussian, xiang2025structured}. At the high level, the current research effort can be broadly categorized as extending the representation~\cite{hamdi2024ges, zhu20253d, liu2025deformable}, designing new optimization strategies~\cite{kheradmand20243d, zhu20253d, kim2025metropolis}, speeding inference processes or/and scaling up the model~\cite{mallick2024taming, kerbl2024hierarchical, feng2025flashgs}. The three lines of research have spawned new applications in novel view synthesis~\cite{yang2025hierarchical, xie2024physgaussian}, SLAM~\cite{matsuki2024gaussian}, geometric reconstruction~\cite{huang20242d, dai2024high}, \etc.

One under-explored theme that is orthogonal to all the afore-mentioned research is data uncertainty, which universally exists in real-world applications. Taking multi-view 4DGS as an example, most existing research focuses on learning the dynamics of Gaussian components, with the aim of high reconstruction quality~\cite{yang2023gs4d, wu20244d}, generalization on unseen motions~\cite{li2024spacetime, zhu2024motiongs}, large complex 3D scenes~\cite{xie2025compositional}, \etc. However, all of them explicitly or implicitly assume that sufficient data can be obtained with high quality. In practice, this often means enough cameras with views covering all angles, and good camera calibration and synchronization. We argue that such assumptions can be too restrictive as the data collection setup can be constrained by many factors such as security, safety, operational constraints, \eg cameras at traffic junctions might only cover a few angles.

Recently, some methods, including contemporaneous ones, have started to consider certain data uncertainties in GS. This includes probabilistic inference for online learning~\cite{van2024variational, savant2024modeling, guo2025uncertainty}, dynamically learning moving and static Gaussians~\cite{gao2024relaygs, wang2025degauss, deng2025dynasplat}, reconstruction from limited camera views~\cite{jeong2024rodygs, jiang2023consistent4d, yilmaz2023107}. However, the data uncertainty explicitly or implicitly considered is mostly specific to one application scenario. Therefore, it is desirable to have one unified framework that can incorporate multiple types of data uncertainty.

We propose Graphical X Splatting (GraphiXS), a new framework which can explicitly incorporate data uncertainty in the form of missing or asynchronized observations in 4DGS. The `X' in GraphiXS is not necessarily Gaussian so we use the word `component'. The term `data uncertainty' broadly refers to missing camera views, sparse camera configurations (\ie position and orientations), missing frames from cameras, imperfectly synchronized cameras, \etc. Probabilistic learning is a natural solution to data uncertainty, but the greatest challenge is to design a flexible probabilistic framework that can model different types of uncertainty in 4DGS. To this end, we formulate GraphiXS as a generative process and propose a new graphical model, by introducing stochasticity into the individual steps of 4DGS. This includes treating all the learnable parameters (\eg component location) as latent variables which are to be inferred via Maximum a Posteriori (MAP). Also, unlike existing methods which treat only the images as observations, we also treat the camera pose and frame time as samples of random variables, which enables us to incorporate the uncertainty in them. Finally, the flexibility of GraphiXS allows us to impose various prior distributions to regulate the behaviors of the components. 

We instantiate GraphiXS with different components including Gaussians and Student's-t to show its generality. Through exhaustive evaluation under various combinations of data uncertainty, we demonstrate that GraphiXS can outperform existing methods across different scenes and metrics. More broadly, we demonstrate that GraphiXS is not a specific model but a framework which can be used to upgrade and improve existing methods. To the best of our knowledge, this is the first probabilistic 4DGS framework targeting multiple types of data uncertainty. Our contributions include: 
\begin{itemize}
    \item A new probabilistic framework to holistically incorporate data uncertainty in the form of sparse spatial and temporal sampling in 4DGS.
    \item A new graphical model which can be instantiated with different primitives and used to `upgrade' existing 4DGS methods.
    \item A new way of introducing stochasticity in the steps of 4DGS.
    \item New priors that can effectively regulate the model behaviors, leading to effective optimization.
\end{itemize}
An open-source implementation of our method is available at {\color{blue}\url{https://github.com/realcrane/Graphical-X-Splatting}}.

\section{Related Work}
\label{sec:related}

\paragraph{Traditional Methods}  Multi-View Stereo (MVS)~\cite{seitz2006comparison} is well studied before deep learning. Software such as COLMAP~\cite{colmap_software} has been widely utilized in 3D reconstruction. It reconstructs the geometrically consistent points between images and restores camera poses by calculating corresponding features in images from multiple perspectives. Denser and more precise geometries can be obtained using the Structure from Motion (SFM)~\cite{ullman1979interpretation}. Meanwhile, additional data (\eg depth) captured from Time of Flight (ToF) or Light Detection And Ranging (Lidar) sensors can be accessed from some other methods such as SLAM~\cite{durrant2006simultaneous, bailey2006simultaneous} and KinectFusion~\cite{newcombe2011kinectfusion} for 3D reconstruction.

\paragraph{Learning-based Reconstruction Method} For a single scene, there are two main cornerstone methods in this sub-area: Neural Radiance Field
(NeRF)~\cite{mildenhall2021nerf} and 3D Gaussian Splatting (3DGS)~\cite{kerbl20233d}. NeRF utilizes a neural network to implicitly learn the 3D radiance field. It can achieve novel view synthesis by querying different points' color and density values from the neural network. NeRF accomplishes good reconstruction results, but its rendering efficiency is low, making it unsuitable for real-time rendering tasks. Comparatively, 3DGS can achieve real-time rendering, but requires more memory compared with NeRF. 3DGS uses 3D Gaussians as components in space. To optimize their attributes (means, covariances, colors, \etc), it uses a rasterization method called Splatting~\cite{zwicker2002ewa} to obtain rendering results from different perspectives. Many subsequent methods are then proposed to improve 3DGS. Among them, some attempt to improve 3DGS in the fundamental paradigm, \eg. using different primitives other than Gaussian including SSS~\cite{zhu20253d}, DBS~\cite{liu2025deformable}, and 2DGS~\cite{huang20242d}, while others try to improve the training processing and adaptive density control in vanilla 3DGS, such as sampling-based~\cite{kheradmand20243d, zhu20253d, kim2025metropolis} methods, elevating rendering quality to a new level.

\paragraph{Dynamic Reconstruction} Most traditional methods either perform a frame-by-frame reconstruction and lack motion continuity, or require calculating optical flow maps to simulate dynamics but cannot reconstruct dense scene flows. In comparison, dynamic reconstruction based on NeRF and 3DGS yields better results. D-NeRF~\cite{pumarola2021d} extend NeRFs to dynamic scenes by introducing an MLP to learn the implicit deformation field in every time interval for the motion. Because of the good learning capability of the implicit representation of NeRF, most NeRF-based dynamic reconstruction methods adopt concepts that are similar to D-NeRF. On the other hand, there are currently two main threads for dynamic reconstruction using 3DGS-based methods. The first is learning the trajectories of explicit primitives at different times. Deformable 3DGS~\cite{yang2024deformable}, 3DGStream~\cite{sun20243dgstream}, and 4DGS-2~\cite{wu20244d} follow the idea of D-NeRF, using neural networks/Tri-planes/HexPlanes to learn the motion and deformation of 3D Gaussians. Building on this, E-D3DGS~\cite{bae2024per} uses per-Gaussian and temporal embeddings for coarse-to-fine motion, while Dynamic 3D Gaussians~\cite{luiten2024dynamic} learn per-frame mean and rotation with fixed appearance parameters. MotionGS~\cite{zhu2024motiongs}, SplineGS~\cite{park2025splinegs}, and FreetimeGS~\cite{wang2025freetimegs}, on the other hand, simulate the dynamics of 3D Gaussians in space through optical flow, spline, and linear flow. The second thread involves building higher-dimensional primitives based on the properties of explicit primitives in the 3DGS method. Then, these properties of primitives at different times in 3D space can be calculated by marginalizing the time dimension. Research such as 4DGS-1~\cite{yang2023gs4d}, 4DRotorGS~\cite{duan20244d}, 7DGS~\cite{gao20257dgs}, and UBS~\cite{liu2025universal} are all under this direction.

\paragraph{Uncertainty in Reconstruction} Existing works such as  SGS~\cite{savant2024modeling} and USPLAT4D~\cite{guo2025uncertainty} have attempted to improve reconstruction quality by estimating the uncertainty of primitives in Gaussian Splatting. However, only few methods take into account the inherent uncertainty of the training data besides the primitive uncertainty. \citet{yang2024deformable} argued that inaccuracies in pose estimation can cause spatial jitter between frames. Research including LongSplat~\cite{lin2025longsplat}, DG-SLAM~\cite{xu2024dg} and GS-CPR~\cite{liu2024gs} adjusts camera poses during the training process. Besides,~\citet{bui2025mobgs} pointed out that the temporal uncertainty should not be neglected, especially for the photon integration process within the physical shutter time. Nevertheless, there is no work modeling further data uncertainty such as asynchronous cameras and performing probabilistic modeling of multiple types of uncertainty for 4DGS.

\section{Methodology}
\label{sec:method}
\subsection{Preliminaries}
\paragraph{3D/4DGS as a mixture model}
GS represents a 3D radiance field using a large set of 3D Gaussians, each parameterized by
\begin{equation}
    \Gaussian(x) = e^{-\frac{1}{2}(x-\mean)^T\cov^{-1}(x-\mean)}
    \label{eq:3DGaussian}
\end{equation}
where $\mean$ and $\Sigma$ are the location and shape (with truncation). In addition, each component has additional attributes, including opacity $\opacity$, and color $\PrimitiveColor$ which is based on spherical harmonics $sh$. During rendering, each 3D Gaussian is transformed into a 2D Gaussian on the image plane. The rendering is then computed by:
\begin{equation}
    C(\pixel) = \sum_{i=1}^{N}\PrimitiveColor_{i}o_{i}\Gaussian^{2D}_{i}(\pixel)\prod_{j=1}^{i-1}(1-o_{j}\Gaussian^{2D}_{j}(\pixel)).
    \label{eq:GS_color}
\end{equation}
where $C(d)$ is the final color of the $d$th pixel, $\NumberOfGaussians$ is the number of the Gaussians that intersect with the ray cast from the pixel. All Gaussian parameters are learned from the 2D images. 

Essentially 3D/4DGS can be seen as learning a (unnormalized) mixture model with Gaussians~\cite{zhu20253d}:
\begin{equation}
    F(x) = \sum w_i \Gaussian_i(x)
    \label{eq:3DGSMixtureModel}
\end{equation}
where $w_i$ is determined by $\opacity_i$, $\PrimitiveColor_i$ and the rendering process. Our GraphiXS generalizes this concept by a new graphical model and a corresponding generative process based on the mixture model. 

\paragraph{Graphical model}
Graphical Model is a probabilistic model which uses a graph to express the conditional dependencies of random variables. It allows structured dependencies to be introduced among random variables to describe a data generation process. Then the knowledge of this process can be used as inductive bias for model design. This is particularly suitable for 4DGS since it describes a multi-step process of rendering (Eq.\ref{eq:3DGaussian}-\ref{eq:3DGSMixtureModel}) and each step involves several quantities which can be treated as random variables. Also, Graphical Model naturally enables Bayesian inference. This is achieved by factorizing the joint probability of all variables into a series of conditional probabilities based on the graph, each of which can be independently modeled and evaluated. For a complex process like 4DGS, as shown later, this provides the flexibility for us to design separate priors to regulate overall model behaviors. 

\subsection{GraphiXS as a Graphical Model}
\label{sec3.2}

\paragraph{Notations} Given videos in a multi-view setting, we aim to reconstruct the 4D radiance field. We define random variables $\CameraPose$ for the cameras pose, $\Times$ for frame time, $\image$ for video frame with pixels $\pixels$. The data consists of $\NumberOfCams$ camera poses, each recording $\NumberOfFrames$ frames, giving a total of $\NumberOfCams \times \NumberOfFrames$ frames and each frame with $\NumberOfPixels$ pixels. We use superscripts for time, superscripts with brackets for derivatives, and subscripts for other indices, \eg $\pixels_{\CurrentCam, \CurrentPixel}^\CurrentTime$ indicates the $\CurrentPixel$th pixel of the image $\image_\CurrentCam^\CurrentTime$ from camera $\CurrentCam$ at time $t$.

\begin{figure}[t]
\vspace{-1em}
    \centering
    \includegraphics[width=0.60\linewidth]{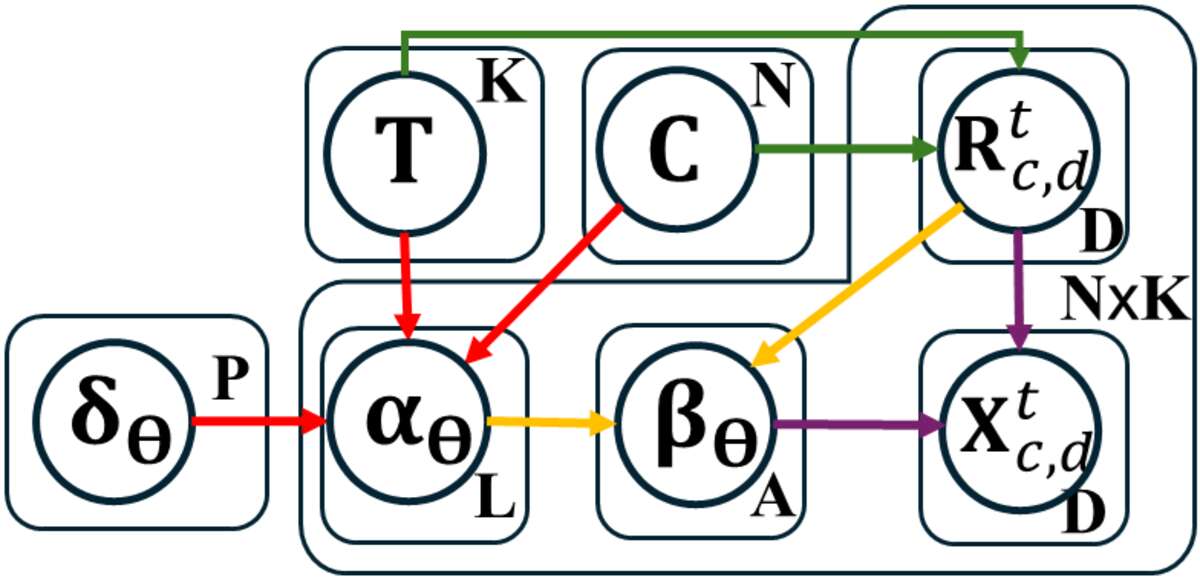}
    \caption{A graphical model for GraphiXS. The colors correspond to the 4 steps of the generative process (1: red, 2: green, 3: yellow, 4: purple)}
    \label{fig:graphicalModel}
\end{figure}
Our new GraphiXS is a graphical model (Fig.~\ref{fig:graphicalModel}) which represents a generative process from a (unnormalized) mixture model with a potentially infinite number of components $F = \sum^{\NumberOfGaussians}_{i=1}\delta_{\theta_i}$, where $\NumberOfGaussians\rightarrow+\infty$. Each component is given by a distribution $\components_{\parameters_i}$ parameterized by $\parameters_i$ (\eg Gaussian, Student’s t-distribution, or other parametric form). The generative process is then described as follows:
\begin{enumerate}
    \item Given the mixture model $F$, a camera $\CurrentCam \sim \CameraPose$ and a time $\CurrentTime \sim \Times$, sample a subset of $F$ which is called $\{\alpha_i\}$ with $\mat{L}$ components, for an image $\image_\CurrentCam^{\CurrentTime}$.
    \item For every $\pixels_{\CurrentCam,\CurrentPixel}^t \in \image_\CurrentCam^\CurrentTime$ indexed by $\CurrentCam$, $\CurrentTime$, and $\CurrentPixel$, sample a ray $\Ray_{\CurrentCam,\CurrentPixel}^t$.
    \item Given $\{\alpha_i\}$ and $\Ray_{c,d}^t$, sample a subset of $\{\alpha_i\}$ which is called $\{\beta_j\}$ with $\mat{A}$ components.
    \item Finally, generate the color for $\pixels_{\CurrentCam,\CurrentTime}^t$ based on $\{\beta_j\}$ and $\Ray_{\CurrentCam,\CurrentPixel}^t$.
\end{enumerate}

Figure~\ref{fig:graphicalModel} describes most of the existing GS work where some steps are realized as rule-based deterministic processes. For instance, the original 3DGS is special case of GraphiXS without $\Times$ and with $\components_{\parameters_i}$ being Gaussian. Then the last two steps correspond to a per-pixel intersection test between a ray and the Gaussians, and rasterization process. The subset $\{\alpha_i\}$ corresponds to the Gaussians that are projected onto the image plane of camera $\CurrentCam$ at time $\CurrentTime$. The subset $\{\beta_j\}$ corresponds to the Gaussians hit by the ray $\Ray_{\CurrentCam,\CurrentPixel}^t$ cast for a pixel.

The key learnable parameter is $\parameters$. It is ideal to make a full Bayesian inference on the posterior distribution of $\parameters$, since there is more than one set of $\parameters$s which can provide good reconstruction. In practice, this is extremely challenging due to the large number of $\parameters$. Models such as Hierarchical Dirichlet Processes could describe our generative process by assuming conjugate priors (\eg Dirichlet-Multinomial), so that intermediate latent variables such as $\alpha$, $\beta$ can be marginalized. However, some steps of the generative process are dictated by light transport which cannot be easily described by probabilistic distributions. Furthermore, even if we forcefully used a full probabilistic model, the optimization would involve intensive sampling or large-scale variational inference~\cite{van2024variational}, both being prohibitively slow in the presence of millions of $\parameters$s. Therefore, we choose Maximum a Posteriori (MAP):
\begin{equation}
    \argmax_{\parameters} P(\pixels, \mid \Ray, \beta, \alpha, \CameraPose, \Times, \components_\parameters) P(\components_\parameters)
    \label{eq:MAP}
\end{equation}
where we assume uninformative priors for $C$ and $T$. Despite not directly modeling their priors, we do implicitly consider their influence on the distribution of the components as random variables, explained later. We first factorize the likelihood following the graphical model:
\begin{equation}
\begin{split}
    P(\pixels \mid \Ray, \beta, &\alpha, \CameraPose, \Times, \components_\parameters) = P(\pixels \mid \bullet)\\
    &=P(\pixels \mid \beta, \Ray)P(\beta \mid \alpha, \Ray)P(\Ray \mid \CameraPose, \Times)P(\alpha \mid \components_\parameters, \CameraPose, \Times),
    \label{eq:likelihood}
\end{split}
\end{equation}
where we use $\bullet$ to represent variables in the condition for brevity. Also, among the possible options for instantiating $\components$ in GraphiXS, \eg Gaussians, Student's-t, Beta, \etc, we assume the component is a distribution parameterized by basic attributes including mean $\mean$, variance $\cov$, color $\PrimitiveColor$, opacity $\opacity$, and other dynamics related attributes introduced later. Although this assumption excludes free-form parameterizations such as shapes~\cite{held20253d}, it is still valid for many GS frameworks. Below, we give the key equations of our instantiations of different distributions in Eq.~(\ref{eq:likelihood}) and provide the details in the supplementary material.

\subsection{Probabilistic Image Reconstruction}
$P(\pixels \mid \bullet)$ represents the image generative process with 4 distributions. For completeness, we explicitly express the full image formation process in the following factorized form with four components. However, several of these steps are dictated by light transport and are inherently deterministic, so we realize them deterministically following existing practice and introduce stochasticity only in the remaining components:
\begin{align}
    P(\pixels \mid \bullet) = \underbrace{P(\pixels \mid \beta, \Ray)}_{\textrm{\textcolor{purple}{rasterization}}}\underbrace{P(\beta \mid \alpha, \Ray)}_{\textrm{\textcolor{brown}{per-pixel comp.}}}
    \underbrace{P(\Ray \mid \CameraPose, \Times)}_{\textrm{\textcolor{teal}{per-pixel ray}}}\underbrace{P(\alpha \mid \components_\parameters, \CameraPose, \Times)}_{\textrm{\textcolor{red}{per-image comp.}}}
    \label{eq:factorizedLik}
\end{align}
Following 3DGS~\cite{kerbl20233d}, we realize $P(\pixels \mid \beta, \Ray)$ as rasterization:
\begin{align}
    Ras(\pixels; \beta, \Ray) = \sum_{i=1}^{N}\lambda_i {\beta_{2D}}_{i}(\pixels), \textrm{ }\lambda_i = \PrimitiveColor_i \opacity_i\prod_{j=1}^{i-1}(1-\opacity_j{\beta_{2D}}_{j}(X))
\end{align}
where $\beta_{2D}$ is a component projected onto the image space.
Similarly, $P(\beta \mid \alpha, \Ray)=Intersect(\Ray, \alpha)$ is the per-pixel intersection test to identify the relevant components $\beta$. $P(\Ray \mid \CameraPose, \Times)=RayCasting(\CameraPose, \Times)$ is ray casting which describes a ray $\Ray$ from a camera $\CurrentCam$ into the space at time $\CurrentTime$. Combining the first three steps gives:
\begin{equation}
X = Ras(\pixels; Intersect(\Ray, \alpha), RayCasting(\CameraPose, \Times)) = Ras(\pixels; \bullet),  
\label{eq:deterministicProcess}
\end{equation}
which is not a distribution and therefore cannot be directly used for MAP. So we use an energy-based distribution~\cite{zhu20253d}:
\begin{align}
    &P(\image \mid \beta, \Ray, \alpha, \CameraPose, \Times) \propto exp(-\sum^\NumberOfCams_c\sum^\NumberOfFrames_t L_{img} (I_c^t)) \nonumber \\
    L_{img} &= (1-\epsilon_{D-SSIM}) L_1 + \epsilon_{D-SSIM} L_{D-SSIM} \nonumber \\
    &+ \epsilon_\opacity \sum_i ||\opacity_i||_1 + \epsilon_{\cov}\sum_i\sum_j ||\sqrt{\lambda_{i,j}}||_1
    \label{eq:imageDis}
\end{align}
where $L_1$ and $L_{D-SSIM}$ are the $L_1$ norm and the structural similarity loss between the reconstructed image (by Eq.~(\ref{eq:deterministicProcess})) and the ground-truth. $\lambda$s are the eigenvalues of $\cov$. The regularization applied to the opacity ensures that the opacity is big only when a component is absolutely needed. The regularization on $\lambda$ ensures the model uses components as spiky as possible (i.e. small variances). Together, the regularization terms minimize the needed number of components.

Next, since there are theoretically an infinite number of configurations of $\parameters$s which can reconstruct the same radiance field, we argue that it is important to impose preferences on well-behaved $\parameters$s. We impose this preference as stochasticity where well-behaved components have higher probabilities, via:
\begin{equation}
    P(\alpha \mid \components_\parameters, \CameraPose, \Times) = E\{P(\CurrentRay^\CurrentTime_{\CurrentCam, \CurrentPixel} | \alpha)\} \approx \frac{ \sum^\NumberOfCams_\CurrentCam\sum^\NumberOfFrames_\CurrentTime \sum^\NumberOfPixels_\CurrentPixel P(\CurrentRay^\CurrentTime_{\CurrentCam, \CurrentPixel} | \alpha)}{\sum^\NumberOfGaussians_p\sum^\NumberOfCams_\CurrentCam\sum^\NumberOfFrames_\CurrentTime \sum^\NumberOfPixels_\CurrentPixel P(\CurrentRay^\CurrentTime_{\CurrentCam, \CurrentPixel} | \components_{\parameters_p})}
    \label{eq:componentConfidence}
\end{equation}
where $P(\CurrentRay^\CurrentTime_{\CurrentCam, \CurrentPixel} | \alpha)$ and $P(\CurrentRay^\CurrentTime_{\CurrentCam, \CurrentPixel} | \components_{\parameters_p})$ are the likelihoods of a ray $\CurrentRay^\CurrentTime_{\CurrentCam, \CurrentPixel}$ from camera $\CurrentCam$ at time $\CurrentTime$ with respect to a component $\alpha$ and $\components_{\parameters_p}$. $\NumberOfCams$, $\NumberOfFrames$, $\NumberOfPixels$, $\NumberOfGaussians$ are the total number of cameras, frames, pixels per image, and components. $P(\CurrentRay^t_{\CurrentCam, \CurrentPixel} | \components_{\parameters_p})$ is the soft visibility of $\components_{\parameters_p}$ to a pixel $x^\CurrentTime_{\CurrentCam, \CurrentPixel}$. When $\components_{\parameters_p}$ has low variance and high $P(\CurrentRay^\CurrentTime_{\CurrentCam, \CurrentPixel} | \components_{\parameters_p})$, $\components_{\parameters_p}$ is highly visible to $\pixels^\CurrentTime_{\CurrentCam, \CurrentPixel}$. Overall, in Eq.~(\ref{eq:componentConfidence}), the numerator is the soft visibility of component $\alpha$ to all pixels of all images across all times, while the denominator is the sum of all the soft visibility of all components. Therefore, we call $P(\alpha \mid \components_{\parameters}, \CameraPose, \Times)$ the component confidence. A component with high confidence is more visible to all pixels in all frames relative to other components.

Aside from the visibility of components to cameras in time, conversely, $P(\alpha \mid \components_{\parameters}, \CameraPose, \Times)$ also implicitly considers the influence of the distributions of $C$ and $T$ on the components. This is important for accommodating sparse views or missing frames. Maximizing this probabilistic distribution means placing components with high probabilities in the overlapped visible regions of multiple cameras and times. If the data is missing from any camera or time, the correctly placed components will be able to make good `guesses' of the 4D radiance field for the missing cameras and frame times, hence robust to these uncertainties.

\subsection{Prior for Component Parameters}
After the likelihood function, we model the prior $P(\components_{\parameters})$ by modeling the distribution of the parameter $\parameters$. We assume our components move in space and time, and parameterize $P(\parameters)$ to capture their dynamics, assuming $\parameters(\CurrentTime)$ is function of $\CurrentTime$:
\begin{equation}
    P(\parameters(\CurrentTime)) = P(\parameters(1)) \prod^{K}_{t=2} P(\parameters(\CurrentTime) \mid \parameters(t-1))
    \label{eq:dynamicsProb}
\end{equation}
The full specification of $\theta=\{ \mean, \cov, \opacity, sh, g, u, v, a, j, s\}$ includes a set of learnable parameters of the components, where $\mean$, $\cov$ are the mean and covariance. $\opacity$ is the base opacity. $sh$ denotes its color represented through spherical harmonics. In time, a component can appear at any time $g$ and last for a period of time $u$. In addition, we also introduce other dynamics related variables $v$, $a$, $j$, $s$ where are detailed later. Similar to~\cite{li2024spacetime}, all parameters in $\theta$ are learned on a per-component basis.

We do not treat all variables in $\parameters$ as functions of time explicitly. Instead, we explicitly model $\mean$ as it is the main variable governing the location of the components. We also make $sh$ dependent on $\mean$. The rest is learned independently. Note that the current $\parameters$ specification is the minimal set of variables to model component dynamics. They are broadly shared in multiple types of components, \eg Gaussian, Beta, Student's-t. There might be other parameters for certain choices of components, such as the control parameter in Student's-t. In this case, these additional parameters are learned independently. 

First, we model $\mean(t)$ as Brownian motions:
\begin{equation}
    \frac{d \mean}{dt} = f(\mean, t) + \mathcal{N}(0, \epsilon^2\mat{I}),
    \label{eq:brownian}
\end{equation}
where $t$ denotes time, $\mathcal{N}$ is a Normal distribution with standard deviation $\epsilon$. $\mat{I}$ is identity matrix. The reason is that we observe that the motions of objects are arbitrary and there is even no guarantee of motion smoothness. Therefore, it is crucial to be able to learn arbitrary and even potentially discontinuous motions. Discretizing Eq.~(\ref{eq:brownian}) in time gives:
\begin{equation}
    \mean(t) = \mean(t-1) + f(\mean, t)\triangle t + \mathcal{N}(0, \epsilon^2\mat{I})\triangle t
    \label{eq:timeDiscretization}
\end{equation}
where the dynamics is governed by $f(\mean, t)$. One simple way of modeling $f(\mean, t)$ is to consider the velocity of the components as in existing methods~\cite{wu20244d}. However, to also consider highly discontinuous motions, we learn several orders of motion derivatives, velocity $v$, acceleration $a$, jerk $j$ and snap $s$:
\begin{equation}
f(\mean, t)\triangle t = v\,(t - g) + \tfrac{1}{2}a\,(t - g)^2 + \tfrac{1}{6}j\,(t - g)^3 + \tfrac{1}{24}s\ (t - g)^4
\label{eq:dynamics}
\end{equation}

Besides, we also design priors on other parameters. This is because MAP for GraphiXS involves millions of parameters. It can easily overfit without prior knowledge to regulate the variables. First, we impose a prior on the base opacity $\opacity$ following~\cite{wang2025freetimegs}:
\begin{equation}
P(o) \propto \frac{1}{\NumberOfGaussians} \sum_{1}^{\NumberOfGaussians} \opacity^2 \exp\!\left(-\frac{1}{2}\left(\frac{t-g}{u}\right)^{2}\right)\,\psi(\cdot),
\label{eq:opacityPrior}
\end{equation}
where $\NumberOfGaussians$ is the total number of components, and $\psi(\cdot)$ denotes the evaluated response of the splatting distribution. The inputs to $\psi(\cdot)$ depend on the chosen component and its associated parameters. For Gaussian, $\psi(\cdot)$ is evaluated using the mean and variance, whereas for Student's-t we additionally include its control parameter $\nu$. Please see the supplementary material for details. Overall this prior penalizes excessively large base opacity values. 

Next, we also impose a prior on $\cov$:
\begin{equation}
    P(\cov_p) \propto exp(-\lambda_{\sigma}\frac{1}{\NumberOfGaussians^t}\sum^{\NumberOfGaussians^t}_p||\cov_p - \hat{\cov^t}||_F^2)
    \label{eq:shapePrior}
\end{equation}
where $\cov_p$ is the covariance of the $p$th component, $\lambda_{\sigma}$ is a weight, $\hat{\cov^t}$ is the mean covariance of components at time $t$, $||\cdot||_F$ is the Frobenius norm, $\NumberOfGaussians^t$ is the total number of components whose temporal support overlaps with time $t$. This prior prefers small shape disparities between components.

We also place a prior on the dynamics related variables:
\begin{equation}
\begin{split}
    P(v_p, a_p, j_p, s_p) \propto exp&(-\lambda_{h}\frac{1}{\NumberOfGaussians^t}\sum^{\NumberOfGaussians^t}_p\sqrt{|\det(\Sigma_p)|}\\ &(||v_p||^2 + ||a_p||^2 + ||j_p||^2 + ||s_p||^2)
    \label{eq:motionPrior}
\end{split}
\end{equation}
which prefers slowly and smoothly moving components. More importantly, it is weighted by the volume proxy of the component $|\det(\cov_p)|$, so that the larger the component is, the slower and smoother its motion should be. This is based on the observation that larger components cover a large area in space. These tend to be in the static background, so they should not move too often and too quickly.

Finally, combining Eq. (\ref{eq:brownian}-\ref{eq:motionPrior}), Eq.~(\ref{eq:dynamicsProb}) becomes:
\begin{equation}
\begin{split}
P&(\parameters(t)) = P(\opacity)P(\mean(1)) \\ &\prod^{K}_{t=2} \mathcal{N}(\mean(t-1) + f(\mean, t), \epsilon^2\mat{I})  \prod^{K}_{t=1}P(v, a, j, s)P(\cov)
\label{eq:componentPrior}
\end{split}
\end{equation}
with the rest parameters learned directly without stochasticity. We learn $P(\parameters(1))$ by maximizing Eq.~(\ref{eq:imageDis}) on the first frame.

\subsection{Objective Function and Optimization}
Finally, with Eq. \ref{eq:imageDis}-\ref{eq:componentPrior}, we have all the elements for MAP (Eq.~(\ref{eq:MAP})):
\begin{align}
    &\argmax_{\parameters} log \ P(\pixels \mid \Ray, \beta, \alpha, \CameraPose, \Times, \parameters) P(\parameters) \nonumber \\
    \Leftrightarrow &\argmin_{\parameters} - log\ [P(\pixels \mid \Ray, \beta, \alpha, \CameraPose, \Times, \parameters)P(\parameters)]
    \label{eq:minNegLik}
\end{align}
where the final loss function is:
\begin{equation}
\begin{split}
    \mathcal{L}_{\textrm{full}}
= \underbrace{\mathcal{L}_{img}}_{\textcolor{red}{-log P(\image \mid \bullet)}}
+ \underbrace{\mathcal{L}_{\alpha}}_{\textcolor{orange}{-log P(\alpha \mid \bullet)}}
+ \underbrace{\mathcal{L}_{\parameters}}_{\textcolor{purple}{-log P(\parameters)}}
    \label{eq:loss}
\end{split}
\end{equation}

For optimization, we use Stochastic Gradient Hamiltonian Monte Carlo (SGHMC)~\cite{zhu20253d}, which injects momentum and controlled stochasticity into gradient updates. In addition, we design component addition/removal sampling and initialization strategies for GraphiXS. Details can be found in the supplementary material.

\section{Experiments}
\paragraph{Datasets} To evaluate GraphiXS, we need to mimic different types of data uncertainty, which requires the original dataset to have enough cameras, views with good coverage, sufficiently high frequency, \etc. Based on the criteria, we choose the Neural 3D Video (N3DV)~\cite{li2022neural} and Google Immersive (GI)~\cite{broxton2020immersive} datasets. Following prior work~\cite{wang2025freetimegs, lee2024fully, wu20244d, yang2023gs4d}, we down-sample all N3DV videos to a resolution of 1352×1014 for both training and evaluation. We consistently hold out the first camera as the test view across all experimental settings and evaluate reconstruction quality on this view over 300 consecutive frames at 30 FPS. Details regarding the GI dataset and additional results are provided in the supplementary material.

\paragraph{Metrics}
We use PSNR (Peak Signal-to-Noise Ratio), DSSIM (Dissimilarity
Structural Similarity Index Measure), and LPIPS (Learned
Perceptual Image Patch Similarity)~\cite{zhang2018unreasonable}. For LPIPS, we report results computed using AlexNet. DSSIM is derived from the multi-scale structural similarity (MS-SSIM) index by converting similarity to a dissimilarity measure and scaled by a factor of 0.5. In all tables, \colorbox{best}{green} and \colorbox{second}{yellow} denote the best and second-best scores, respectively.

\paragraph{Baseline Methods}
We choose a representative set of state-of-the-art methods as baselines including 4DGS-1~\cite{yang2023gs4d}, 4DGS-2~\cite{wu20244d}, Ex4DGS~\cite{lee2024fully}, and FreeTimeGS~\cite{wang2025freetimegs}. We use their official open-source implementations when available, otherwise our own implementation (for FreeTimeGS~\cite{wang2025freetimegs}). For fair comparison, we train all models from scratch and run them 3 times and report the average. We color the best and second best results in green and yellow respectively. We report averaged metrics across scenes in the main paper and provide detailed results and analysis in the supplementary material.

\paragraph{Uncertainty Settings} We design several settings for multiple types of commonly seen data uncertainty. Standard Setting uses exactly the same setting as the baseline methods, regarded as without any data uncertainty. Sparse Views represents missing cameras. Sparse Frames represents only low frequency cameras are used. Unsynchronized Cameras represents a group of cameras which are not synchronized. Faulty Cameras represents a system with random camera malfunctions. Together these settings cover a wide range of possible scenarios where data uncertainty is induced into the data.

\paragraph{Instantiation and Generalization} GraphiXS does not assume specific components, so we instantiate it with two components to show its generality. The first is Student's-t  (GraphiTS) and the second is approximate Gaussian (GraphiGS) by fixing the control parameter of Student's-t to a large value. In addition, we show that GraphiXS can be used to `upgrade' existing 4DGS methods.

\subsection{Comparison under Standard Setting}
We show the numerical results for both N3DV and GI datasets in \cref{tab:results_standard_avg}. Overall, the DSSIM for most methods are similar showing all methods capture the structure of images well in reconstruction. Both GraphiTS and GraphiGS outperform existing methods, mainly on PSNR and LPIPS. This is somewhat surprising, as the camera angles in N3DV are dense \eg little data uncertainty, so one would assume further modeling of data uncertainty is redundant. However, the experiments demonstrate that even with the full observations, explicitly considering data uncertainty can further enhance the reconstruction quality. This becomes more pronounced on the GI dataset, where the standard camera setup is sparser compared to N3DV. We show one visual result in \cref{fig:dynamics_model_comparison}. This is a difficult scene. One example is that one hand of the person holding a tong moves quickly from time to time, causing visual blur. This motion involves constantly stirring the spinach in random manner both spatially and in terms of its dynamics, causing motion blur. So capturing the high-order dynamics is crucial. Therefore, comparatively GraphiXS reconstructs clearer structures of the hand and tong with details, demonstrating the dynamics of the components are learned well. 

\begin{table}[t!]
\caption{Comparison under standard setting on N3DV and GI datasets.}
\label{tab:results_standard_avg}
\centering
\scriptsize
\setlength{\tabcolsep}{0.6pt}
\renewcommand{\arraystretch}{1.05}

\begin{tabular}{l|ccc|ccc}
\toprule
\multirow{2}{*}[-1.0ex]{\textbf{Method}}
& \multicolumn{3}{c|}{\textbf{N3DV}}
& \multicolumn{3}{c}{\textbf{GI}} \\
\cmidrule(lr){2-4} \cmidrule(lr){5-7}
& \begin{tabular}[c]{@{}c@{}}\textbf{PSNR}$\uparrow$\end{tabular}
& \begin{tabular}[c]{@{}c@{}}\textbf{DSSIM}$\downarrow$\end{tabular}
& \begin{tabular}[c]{@{}c@{}}\textbf{LPIPS}$\downarrow$\end{tabular}
& \begin{tabular}[c]{@{}c@{}}\textbf{PSNR}$\uparrow$\end{tabular}
& \begin{tabular}[c]{@{}c@{}}\textbf{DSSIM}$\downarrow$\end{tabular}
& \begin{tabular}[c]{@{}c@{}}\textbf{LPIPS}$\downarrow$\end{tabular} \\
\midrule

4DGS-1
& 31.18 & \cellcolor{best}0.015 & 0.051
& 26.76 & 0.030 & 0.153 \\

4DGS-2
& 30.52 & 0.019 & 0.060
& 23.99 & 0.079 & 0.290 \\

Ex4DGS
& 31.71 & \cellcolor{best}0.015 & 0.050
& 20.39 & 0.107 & 0.190 \\

FTGS
& 31.55 & \cellcolor{second}0.016 & 0.047
& 28.01 & 0.022 & 0.073 \\

\midrule
\textbf{Ours (GraphiGS)}
& \cellcolor{second}31.78 & \cellcolor{best}0.015 & \cellcolor{second}0.044
& \cellcolor{second}28.84 & \cellcolor{second}0.017 & \cellcolor{second}0.064 \\

\textbf{Ours (GraphiTS)}
& \cellcolor{best}32.02 & \cellcolor{best}0.015 & \cellcolor{best}0.043
& \cellcolor{best}29.08 & \cellcolor{best}0.016 & \cellcolor{best}0.056 \\

\bottomrule
\end{tabular}
\end{table}

\subsection{Comparison under Sparse Views}
In Sparse Views, we randomly remove 10\%, 30\%, and 50\% of the training cameras, causing view gaps. \Cref{tab:results_spatial_sparsity_avg} shows the numerical comparison on N3DV dataset. Overall, GraphiXS outperforms or is in par with other methods across all metrics. Due to the dense nature of the cameras in N3DV, a 10\% reduction of the cameras does not challenge the methods, sometimes even improve the results (\eg PSNR in 4DGS-1, 4DGS-2 and Ex4DGS). Further analysis suggests that the difference might be due to the randomness in the training for different methods, which might cause bigger variances in different runs in some methods but overall give similar results to the Standard Setting. When dropping 30\% and 50\% of the cameras, all metrics start to deteriorate, showing Sparse Views induces major data uncertainty. Across all three settings, GraphiXS achieves the best in 8 of 9 experiments and the second best in 1 experiment. We show one visual result in \cref{fig:teaser} Left and more results in \cref{fig:spatial_sparse_fig}. When more cameras are missing, it becomes more challenging for all methods where GraphiXS is affected the least.

\begin{table}[t!]
\caption{Comparison at different levels of spatial sparsity.}
\label{tab:results_spatial_sparsity_avg}
\centering
\scriptsize
\setlength{\tabcolsep}{0.60pt}
\renewcommand{\arraystretch}{1.05}

\begin{tabular}{l|ccc|ccc|ccc}
\toprule
\multirow{2}{*}[-1.0ex]{\textbf{Method}}
& \multicolumn{3}{c}{\textbf{10\%}}
& \multicolumn{3}{c}{\textbf{30\%}}
& \multicolumn{3}{c}{\textbf{50}\%} \\
\cmidrule(lr){2-4} \cmidrule(lr){5-7} \cmidrule(lr){8-10}
& \begin{tabular}[c]{@{}c@{}}\textbf{PSNR}$\uparrow$\end{tabular}
& \begin{tabular}[c]{@{}c@{}}\textbf{DSSIM}$\downarrow$\end{tabular}
& \begin{tabular}[c]{@{}c@{}}\textbf{LPIPS}$\downarrow$\end{tabular}
& \begin{tabular}[c]{@{}c@{}}\textbf{PSNR}$\uparrow$\end{tabular}
& \begin{tabular}[c]{@{}c@{}}\textbf{DSSIM}$\downarrow$\end{tabular}
& \begin{tabular}[c]{@{}c@{}}\textbf{LPIPS}$\downarrow$\end{tabular}
& \begin{tabular}[c]{@{}c@{}}\textbf{PSNR}$\uparrow$\end{tabular}
& \begin{tabular}[c]{@{}c@{}}\textbf{DSSIM}$\downarrow$\end{tabular}
& \begin{tabular}[c]{@{}c@{}}\textbf{LPIPS}$\downarrow$\end{tabular} \\
\midrule
4DGS-1
& 31.37 & \cellcolor{best}0.015 & 0.049
& 29.11 & \cellcolor{best}0.024 & \cellcolor{best}0.058
& 28.54 & \cellcolor{second}0.027 & 0.068 \\

4DGS-2
& 31.00 & \cellcolor{second}0.016 & 0.057
& 28.68 & \cellcolor{best}0.024 & 0.069
& 27.99 & 0.029 & 0.074 \\

Ex4DGS
& 30.94 & \cellcolor{second}0.016 & 0.051
& 28.86 & 0.026 & 0.064
& 28.33 & 0.028 & 0.067 \\

FTGS
& 31.38 & \cellcolor{second}0.016 & 0.047
& 29.04 & 0.026 & 0.063
& 28.06 & 0.031 & 0.072 \\

\midrule
\textbf{Ours (GraphiGS)}
& \cellcolor{best}31.73 & \cellcolor{second}0.016 & \cellcolor{best}0.044
& \cellcolor{second}29.35 & \cellcolor{second}0.025 & \cellcolor{best}0.058
& \cellcolor{best}28.60 & \cellcolor{best}0.026 & \cellcolor{best}0.066 \\

\textbf{Ours (GraphiTS)}
& \cellcolor{second}31.62 & \cellcolor{second}0.016 & \cellcolor{second}0.046
& \cellcolor{best}29.41 & \cellcolor{best}0.024 & \cellcolor{second}0.061
& \cellcolor{second}28.56 & \cellcolor{second}0.027 & \cellcolor{second}0.067 \\
\bottomrule
\end{tabular}
\end{table}

\subsection{Comparison under Sparse Frames}
In Sparse Frames, we reduce the frame rate of the training cameras from 30 FPS to 20 and 10 FPS while still evaluating the scene at 30 FPS causing time sparsity. 
\Cref{tab:results_temporal_sparsity_avg} shows the numerical results for both settings on N3DV dataset. Overall GraphiXS outperforms other methods. Different from Sparse Views, Sparse Frames mimics low frequency cameras which are not suitable for recording fast motions. Since motions in N3DV are generally not fast, and all methods have dedicated parts to learn the dynamics of the components, Sparse Frames is generally less challenging than Sparse Views. 

Furthermore, the key difference between different methods stem from how dense in time they require samples to be. However individual model behaviors do not share the same pattern. For the baseline methods, when FPS is lower, the results become worse as expected. The only exception is 4DGS-2 which is slightly improved. Our speculation is that since 4DGS-2 exhaustively learns pair-wise relationships between the $x$, $y$, $z$ coordinates of the Gaussian mean and time $t$, there is a chance their network overfits when the sample density in time is too high. So down-sampling actually improves the results. Last, GraphiGS does not deteriorate from Standard Setting, GraphiTS deteriorates at 20 FPS but not further in 10 FPS. We show one visual result in \cref{fig:teaser} Middle and more results in \cref{fig:temporal_sparse_fig}.

\begin{table}[tb]
\caption{Comparison at different levels of temporal sparsity.}
\label{tab:results_temporal_sparsity_avg}
\centering
\scriptsize
\setlength{\tabcolsep}{2.8pt}
\renewcommand{\arraystretch}{1.05}

\begin{tabular}{l|ccc|ccc}
\toprule
\multirow{2}{*}[-1.0ex]{\textbf{Method}}
& \multicolumn{3}{c}{\textbf{20 FPS}}
& \multicolumn{3}{c}{\textbf{10 FPS}} \\
\cmidrule(lr){2-4} \cmidrule(lr){5-7}
& \begin{tabular}[c]{@{}c@{}}\textbf{PSNR}$\uparrow$\end{tabular}
& \begin{tabular}[c]{@{}c@{}}\textbf{DSSIM}$\downarrow$\end{tabular}
& \begin{tabular}[c]{@{}c@{}}\textbf{LPIPS}$\downarrow$\end{tabular}
& \begin{tabular}[c]{@{}c@{}}\textbf{PSNR}$\uparrow$\end{tabular}
& \begin{tabular}[c]{@{}c@{}}\textbf{DSSIM}$\downarrow$\end{tabular}
& \begin{tabular}[c]{@{}c@{}}\textbf{LPIPS}$\downarrow$\end{tabular} \\
\midrule
4DGS-1
& 31.26 & \cellcolor{best}0.015 & 0.049
& 30.72 & \cellcolor{second}0.016 & 0.051 \\

4DGS-2
& 30.46 & \cellcolor{second}0.016 & 0.058
& 31.01 & \cellcolor{second}0.016 & 0.059 \\

Ex4DGS
& 31.39 & \cellcolor{best}0.015 & 0.050
& 31.15 & \cellcolor{second}0.016 & 0.053 \\

FTGS
& 31.39 & 0.017 & \cellcolor{second}0.046
& 31.36 & 0.017 & \cellcolor{second}0.046 \\

\midrule
\textbf{Ours (GraphiGS)}
& \cellcolor{best}31.88 & \cellcolor{best}0.015 & \cellcolor{best}0.043
& \cellcolor{best}31.87 & \cellcolor{best}0.015 & \cellcolor{best}0.044 \\

\textbf{Ours (GraphiTS)}
& \cellcolor{second}31.62 & \cellcolor{best}0.015 & \cellcolor{best}0.043
& \cellcolor{second}31.63 & \cellcolor{best}0.015 & \cellcolor{best}0.044 \\
\bottomrule
\end{tabular}
\end{table}

\subsection{Comparison under Unsynchronized Cameras}
In data capture, synchronizing cameras require extra hardware, software and calibration effort. So we test if methods could work with unsynchronized cameras. We simulate this scenario by setting cameras at different FPS. We randomly select 10\% and 50\% of cameras and reduce their FPS from 30 to 20, yielding 2 settings with two levels of partial temporal sparsity. \Cref{tab:results_nonsync_temporal_sparsity_avg} shows the results on N3DV dataset. Overall, we find Unsynchronized Cameras is an easier setting for all methods compared with Sparse Views and Sparse Frames. This is expected as all methods learn the dynamics of the components which does not require observations to be available for arbitrary time. Unsynchronized Cameras could be a bigger issue if cameras are not dense in space where every frame from every camera is crucial in capturing the motions. But in N3DV, all the cameras are in front of the person. Overall, GraphiXS gives the best results.

\begin{table}[tb]
\caption{Comparison under different levels of Unsynchronization.}
\label{tab:results_nonsync_temporal_sparsity_avg}
\centering
\scriptsize
\setlength{\tabcolsep}{2.8pt}
\renewcommand{\arraystretch}{1.05}

\begin{tabular}{l|ccc|ccc}
\toprule
\multirow{2}{*}[-1.0ex]{\textbf{Method}}
& \multicolumn{3}{c}{\textbf{10\% @ 20 FPS}}
& \multicolumn{3}{c}{\textbf{50\% @ 20 FPS}} \\
\cmidrule(lr){2-4} \cmidrule(lr){5-7}
& \begin{tabular}[c]{@{}c@{}}\textbf{PSNR}$\uparrow$\end{tabular}
& \begin{tabular}[c]{@{}c@{}}\textbf{DSSIM}$\downarrow$\end{tabular}
& \begin{tabular}[c]{@{}c@{}}\textbf{LPIPS}$\downarrow$\end{tabular}
& \begin{tabular}[c]{@{}c@{}}\textbf{PSNR}$\uparrow$\end{tabular}
& \begin{tabular}[c]{@{}c@{}}\textbf{DSSIM}$\downarrow$\end{tabular}
& \begin{tabular}[c]{@{}c@{}}\textbf{LPIPS}$\downarrow$\end{tabular} \\
\midrule
4DGS-1
& 31.57 & \cellcolor{second}0.015 & 0.049
& 31.54 & \cellcolor{best}0.015 & 0.050 \\

4DGS-2
& 30.68 & 0.016 & 0.059
& 30.92 & \cellcolor{second}0.016 & 0.059 \\

Ex4DGS
& 31.34 & 0.016 & 0.050
& 31.01 & \cellcolor{second}0.016 & 0.051 \\

FTGS
& 31.44 & 0.017 & 0.047
& 31.34 & 0.017 & 0.047 \\

\midrule
\textbf{Ours (GraphiGS)}
& \cellcolor{second}31.78 & \cellcolor{second}0.015 & \cellcolor{second}0.044
& \cellcolor{second}31.76 & \cellcolor{best}0.015 & \cellcolor{second}0.044 \\

\textbf{Ours (GraphiTS)}
& \cellcolor{best}31.87 & \cellcolor{best}0.014 & \cellcolor{best}0.043
& \cellcolor{best}31.85 & \cellcolor{best}0.015 & \cellcolor{best}0.043 \\
\bottomrule
\end{tabular}
\end{table}

\subsection{Comparison under Faulty Cameras}
During data recording, any camera can malfunction which will cause a combination of all the types of the data uncertainty before this section. This data uncertainty spans across space and time. We design three settings to simulate Faulty Camera. These settings aggregate the previous parameters, resulting in total space-time sparsities of approximately 13\% for Faulty Cam 1, 37\% for Faulty Cam 2 and 84\% for Faulty Cam 3. \Cref{tab:results_cam_fail_avg} shows the numerical results for these settings on N3DV dataset. One visual example is shown in \cref{fig:teaser} Right and more are in \cref{fig:faulty_cam_fig}.

\begin{table}[tb]
\caption{Comparison with Faulty Camera settings.}
\label{tab:results_cam_fail_avg}
\centering
\scriptsize
\setlength{\tabcolsep}{0.60pt}
\renewcommand{\arraystretch}{1.05}

\begin{tabular}{l|ccc|ccc|ccc}
\toprule
\multirow{2}{*}[-1.0ex]{\textbf{Method}}
& \multicolumn{3}{c}{\textbf{Faulty Cam 1}}
& \multicolumn{3}{c}{\textbf{Faulty Cam 2}}
& \multicolumn{3}{c}{\textbf{Faulty Cam 3}} \\
\cmidrule(lr){2-4} \cmidrule(lr){5-7} \cmidrule(lr){8-10}
& \begin{tabular}[c]{@{}c@{}}\textbf{PSNR}$\uparrow$\end{tabular}
& \begin{tabular}[c]{@{}c@{}}\textbf{DSSIM}$\downarrow$\end{tabular}
& \begin{tabular}[c]{@{}c@{}}\textbf{LPIPS}$\downarrow$\end{tabular}
& \begin{tabular}[c]{@{}c@{}}\textbf{PSNR}$\uparrow$\end{tabular}
& \begin{tabular}[c]{@{}c@{}}\textbf{DSSIM}$\downarrow$\end{tabular}
& \begin{tabular}[c]{@{}c@{}}\textbf{LPIPS}$\downarrow$\end{tabular}
& \begin{tabular}[c]{@{}c@{}}\textbf{PSNR}$\uparrow$\end{tabular}
& \begin{tabular}[c]{@{}c@{}}\textbf{DSSIM}$\downarrow$\end{tabular}
& \begin{tabular}[c]{@{}c@{}}\textbf{LPIPS}$\downarrow$\end{tabular} \\
\midrule

4DGS-1
& 30.77 & \cellcolor{best}0.017 & 0.051 
& \cellcolor{second}30.31 & \cellcolor{best}0.020 & 0.055
& 19.29 & 0.136 & 0.314 \\

4DGS-2
& 30.56 & \cellcolor{second}0.018 & 0.061 
& 29.78 & \cellcolor{second}0.021 & 0.065
& 14.93 & 0.202 & 0.491 \\

Ex4DGS
& 30.52 & 0.019 & 0.053 
& 30.03 & \cellcolor{best}0.020 & 0.056
& 16.38 & 0.203 & 0.359 \\

FTGS
& 30.46 & 0.020 & 0.051 
& 29.90 & 0.026 & \cellcolor{best}0.051
& 20.49 & \cellcolor{best}0.115 & \cellcolor{best}0.228 \\

\midrule

\textbf{Ours (GraphiGS)}
& \cellcolor{best}30.90 & \cellcolor{second}0.018 & \cellcolor{best}0.048
& \cellcolor{best}30.39 & \cellcolor{best}0.020 & \cellcolor{second}0.052
& \cellcolor{second}20.65 & \cellcolor{second}0.118 & \cellcolor{second}0.250 \\

\textbf{Ours (GraphiTS)}
& \cellcolor{second}30.79 & 0.019 & \cellcolor{second}0.050 
& 30.18 & \cellcolor{second}0.021 & 0.053
& \cellcolor{best}20.82 & 0.119 & 0.253 \\

\bottomrule
\end{tabular}
\end{table}

\subsection{GraphiXS as an Upgrade}
GraphiXS is not only a specific method, but a framework that can be used to `upgrade' other compatible methods. This involves turning their deterministic models into probabilistic ones, using formulations similar to \cref{eq:imageDis}, then modeling stochasticity. The former step varies depending on the specific method, and the latter step is to add our stochastic components such as $P(\alpha \mid \theta, C, T)$ and $P(\Sigma_p)$. We directly show results here and give the details in the supplementary material. We use the Standard Setting and the Faulty Camera 1 and 2 settings as it includes all types of data uncertainty. Among the latest methods, we choose FTGS~\cite{wang2025freetimegs} and \Cref{tab:upgrade} shows the numerical results. One visual example is shown in \cref{fig:upgrade_fig}.

\begin{table}[tb]
\caption{Comparison in before and after upgrade for FTGS across Standard and Faulty Camera settings.}
\label{tab:upgrade}
\scriptsize
\setlength{\tabcolsep}{0.60pt}
\renewcommand{\arraystretch}{1.05}

\begin{tabular}{l|ccc|ccc|ccc}
\toprule
\multirow{2}{*}[-1.0ex]{\textbf{Method}}
& \multicolumn{3}{c}{\textbf{Standard}}
& \multicolumn{3}{c}{\textbf{Faulty Cam 1}}
& \multicolumn{3}{c}{\textbf{Faulty Cam 2}} \\
\cmidrule(lr){2-4} \cmidrule(lr){5-7} \cmidrule(lr){8-10}
& \begin{tabular}[c]{@{}c@{}}\textbf{PSNR}$\uparrow$\end{tabular}
& \begin{tabular}[c]{@{}c@{}}\textbf{DSSIM}$\downarrow$\end{tabular}
& \begin{tabular}[c]{@{}c@{}}\textbf{LPIPS}$\downarrow$\end{tabular}
& \begin{tabular}[c]{@{}c@{}}\textbf{PSNR}$\uparrow$\end{tabular}
& \begin{tabular}[c]{@{}c@{}}\textbf{DSSIM}$\downarrow$\end{tabular}
& \begin{tabular}[c]{@{}c@{}}\textbf{LPIPS}$\downarrow$\end{tabular}
& \begin{tabular}[c]{@{}c@{}}\textbf{PSNR}$\uparrow$\end{tabular}
& \begin{tabular}[c]{@{}c@{}}\textbf{DSSIM}$\downarrow$\end{tabular}
& \begin{tabular}[c]{@{}c@{}}\textbf{LPIPS}$\downarrow$\end{tabular} \\
\midrule

FTGS
& 31.55 & 0.016 & 0.047
& 30.46 & 0.020 & 0.051
& 29.90 & 0.026 & 0.051 \\

FTGS UG
& 31.61 & 0.016 & 0.044
& 30.80 & 0.019 & 0.050
& 30.20 & 0.021 & 0.054 \\

\bottomrule
\end{tabular}
\end{table}

\subsection{Ablation Study}
Since we decompose GraphiXS into distributions (\cref{eq:likelihood}) and propose various parameterization for each distribution, we show their respective effectiveness. The key distributions include the Higher Order Dynamics (\cref{eq:brownian}) and the component confidence $P(\alpha \mid \components_\parameters, \CameraPose, \Times)$(\cref{eq:componentConfidence}). We also include a deterministic variant of GraphiXS to assess the benefit of our probabilistic formulation (see the supplementary material for details). Therefore, we conduct an ablation study on GraphiGS under the three settings on Faulty Cam 1 and 2. We show the quantitative results in \Cref{tab:ablation} and qualitative results in \cref{fig:ablation_fig}. In general, removing any component will deteriorate the results. This is more obvious in Faulty Cams 2 which has a higher level of uncertainty. `W/O Higher Order Dynamics' only considers the position and velocity of the component, which is a strategy for many existing 4DGS methods~\cite{wang2025freetimegs, lee2024fully, li2024spacetime}. But the results shows that considering higher order dynamics will improve the results. Furthermore, `W/O $P(\alpha \mid \theta, C, T)$' shows that the component confidence part also improves the results. This part is designed to regulate component locations and shapes to make them visible to cameras in time. 
Lastly, the deterministic variant shows a significant performance drop, demonstrating the effectiveness of our framework.

\begin{table}[tb]
\caption{Quantitative results of our ablation study on GraphiGS.}
\label{tab:ablation}
\centering
\scriptsize
\setlength{\tabcolsep}{2.8pt}
\renewcommand{\arraystretch}{1.05}

\begin{tabular}{l|ccc|ccc}
\toprule
\multirow{2}{*}[-1.0ex]{\textbf{Method}}
& \multicolumn{3}{c}{\textbf{Faulty Cam 1}}
& \multicolumn{3}{c}{\textbf{Faulty Cam 2}} \\
\cmidrule(lr){2-4} \cmidrule(lr){5-7}
& \begin{tabular}[c]{@{}c@{}}\textbf{PSNR}$\uparrow$\end{tabular}
& \begin{tabular}[c]{@{}c@{}}\textbf{DSSIM}$\downarrow$\end{tabular}
& \begin{tabular}[c]{@{}c@{}}\textbf{LPIPS}$\downarrow$\end{tabular}
& \begin{tabular}[c]{@{}c@{}}\textbf{PSNR}$\uparrow$\end{tabular}
& \begin{tabular}[c]{@{}c@{}}\textbf{DSSIM}$\downarrow$\end{tabular}
& \begin{tabular}[c]{@{}c@{}}\textbf{LPIPS}$\downarrow$\end{tabular} \\
\midrule

W/O Higher Order Dynamics
& 30.84 & 0.019 & 0.049
& 30.03 & 0.021 & 0.054 \\

W/O $P(\alpha \mid \theta, C, T)$
& 30.74 & 0.019 & 0.051
& 30.04 & 0.022 & 0.054 \\

W/O Stochasticity
& 28.64 & 0.030 & 0.100
& 27.66 & 0.036 & 0.113 \\

\textbf{Ours (GraphiGS)}
& 30.90 & 0.018 & 0.048
& 30.39 & 0.020 & 0.052 \\

\bottomrule
\end{tabular}
\end{table}

\begin{figure*}[!b]
    \centering

    \figrowCrops{figs/dynamics_model_comparison}
    \caption{
        Comparison using N3DV dataset under standard setting. Per-region PSNR scores are given at the top right of each image. Enlarged regions contain complex and fast motions \eg the tongs motion. GraphiXS reconstruction is more clear and richer in details than other methods.
    }
    \label{fig:dynamics_model_comparison}
    \rule{0pt}{2em}
    
    \figrowCropsThreeRowWL{figs/spatial_sparse_fig}{sparse_90}{sparse_70}{sparse_50}{10\%}{30\%}{50\%}
    \caption{
        Comparison using GI dataset under 10\%, 30\%, and 50\% spatial sparsity. Per-region PSNR scores are given at the top right of each crop. GraphiXS is affected the least when the percentage of missing cameras increases.
    }
    \label{fig:spatial_sparse_fig}
\end{figure*}

\section{Conclusion and Discussion}
We have proposed the first probabilistic 4DGS framework, GraphiXS, that holistically considers multiple types of data uncertainty. GraphiXS is general in that it can be instantiated with different components or used to upgrade existing methods. Through exhaustive evaluation, we have demonstrated the effectiveness of GraphiXS. A major limitation is GraphiXS assumes the components are probabilistic distributions parameterized by a set of common parameters. This makes it unsuitable for upgrading methods with other types of components such as geometric primitives. Also, GraphiXS does not do full Bayesian inference, \ie no posterior distribution learned for $\parameters$, which would be ideal given the stochastic nature of 4DGS models. Lastly, GraphiXS assumes uninformative priors for $C$ and $T$, and could be extended to incorporate camera pose priors $P(C)$; we leave this to future work.
\balance

\begin{acks}
This work was supported in part by the Dr. Rabin Ezra Scholarship (Charity No. 1116049), awarded to Do\u{g}a Y\i lmaz, and the UK Research and Innovation AIRR Innovator Award (0261-5654-9320-1).
\end{acks}

\begin{figure*}[p]
    \centering

    \figrowCropsTwoRowWL{figs/temporal_sparse_fig}{fps_20}{fps_10}{20 FPS}{10 FPS}
    \caption{
        Comparison using N3DV dataset under 20 FPS and 10 FPS temporal sparsity. Per-region PSNR scores are given at the top right of each crop. GraphiXS is affected the least when the training camera FPS drops.
    }
    \label{fig:temporal_sparse_fig}

    \rule{0pt}{2.5em} 

    \figrowCropsThreeRowWL{figs/faulty_cam_fig}{faulty_cams}{faulty_cams_2}{faulty_cams_3}{Faulty Cam 1}{Faulty Cam 2}{Faulty Cam 3}
    \caption{
        Comparison using GI dataset under faulty camera settings. Per-region PSNR scores are given at the top right of each crop. GraphiXS is affected the least when spatio-temporal sparsity is increased.
    }
    \label{fig:faulty_cam_fig}

    \rule{0pt}{2.5em} 

    \figrowCropsTwoRowWLShort{figs/upgrade_fig}{fcam}{fcam2}{Faulty Cam 1}{Faulty Cam 2}
    \caption{
        Comparison of FTGS~\cite{wang2025freetimegs} with and without upgrading under faulty camera 1 and 2 settings. Per-region PSNR scores are given at the top right of each crop. Upgrading FTGS using GraphiXS improves visual quality under various levels of spatio-temporal uncertainty.
    }
    \label{fig:upgrade_fig}
\end{figure*}

\clearpage

\begin{figure*}[!t]
    \centering

    \figrowCropsTwoRowWLShortAbl{figs/ablation_fig}{fcam}{fcam2}{Faulty Cam 1}{Faulty Cam 2}
    \caption{
        Visual results of our ablation study. Per-region PSNR scores are given at the top right of each crop.
    }
    \label{fig:ablation_fig}
\end{figure*}

\putbib[main]
\end{bibunit}

\clearpage
\begin{bibunit}[ACM-Reference-Format]
\input{supp}
\putbib[main]
\end{bibunit}

\end{document}

%% file: supp.tex
\clearpage

\setcounter{page}{1}

\setcounter{section}{0}
\setcounter{figure}{0}
\setcounter{table}{0}
\setcounter{equation}{0}

\renewcommand{\thesection}{\arabic{section}}
\renewcommand{\thefigure}{\arabic{figure}}
\renewcommand{\thetable}{\arabic{table}}
\renewcommand{\theequation}{\arabic{equation}}

\makeatletter
\gdef\@abstract{}
\gdef\@teaserfigures{}
\makeatother

\title[GraphiXS Supplementary Material]{Graphical X Splatting (GraphiXS): A Graphical Model for 4D Gaussian Splatting under Uncertainty Supplementary Material}

\maketitle

\section{Additional Experimental Results}

\subsection{Quantitative Results}
We report comprehensive per-scene quantitative metrics on the Neural 3D Video (N3DV)~\cite{li2022neural} dataset in \cref{tab:results_standard_full,tab:results_spatial_sparsity_full,tab:results_sync_temporal_sparsity_full,tab:results_nonsync_temporal_sparsity_full,tab:results_fcam_full}. We further present analogous results on the Google Immersive (GI)~\cite{broxton2020immersive} dataset in \cref{tab:results_standard_full_gi,tab:results_spatial_sparsity_full_gi,tab:results_sync_temporal_sparsity_full_gi,tab:results_nonsync_temporal_sparsity_full_gi,tab:results_fcam_full_gi}. In addition, we include results on a synthetic scene where we have full control over camera configurations in \cref{tab:results_citycrowd}. In all tables, \colorbox{best}{green} and \colorbox{second}{yellow} denote the best and second-best scores, respectively. Overall, these results are consistent with the average metrics reported in the main manuscript. However, the Coffee Martini scene exhibits a trend that differs from the other sequences. This discrepancy is caused by a known temporal synchronization artifact in Camera 13 of the Coffee Martini scene. Camera 13 is temporally misaligned with respect to the other cameras, leading to consistent frame-level offsets between its observations and the true scene dynamics. As a result, images from Camera 13 correspond to different time instants than those assumed by methods that rely on strict multi-view temporal alignment. This issue primarily affects approaches that explicitly model dynamic geometry and appearance from synchronized multi-view inputs. In particular, GraphiXS and other explicit dynamic reconstruction methods assume consistent temporal correspondence across cameras and therefore interpret the desynchronized observations as conflicting geometric or appearance changes, which degrades reconstruction quality and quantitative performance. In contrast, 4DGS-2 outperforms GraphiXS and other baselines in this setting. We attribute this behavior to the implicit nature of the 4DGS-2 representation, which makes it more tolerant to moderate temporal misalignment in the input camera streams. As a result, 4DGS-2 is less affected by the Camera 13 synchronization error, leading to improved quantitative performance on the Coffee Martini scene despite the underlying dataset error. This effect is also reflected in the Coffee Martini columns of \cref{tab:results_standard_full,tab:results_spatial_sparsity_full,tab:results_sync_temporal_sparsity_full,tab:results_nonsync_temporal_sparsity_full,tab:results_fcam_full}, where 4DGS-2 achieves the highest PSNR in seven out of ten evaluated settings.

For the GI dataset, we evaluate on a representative subset of scenes: Welder, Flames, Horse, Goats, and Alexa10. These scenes provide diverse challenges, including fast-moving particles (Welder, Flames), outdoor environments (Horse, Goats), and complex color patterns (Alexa10). For all scenes, we designate the first camera as the test view and evaluate over 30 frames at quarter resolution. 
The GI capture setup employs outward-facing cameras on a spherical rig, in contrast to N3DV where cameras are inward-facing toward the subject. This results in reduced view overlap and a more spatially sparse capture regime. Under these conditions, the performance gap between GraphiXS and competing baselines becomes more pronounced, highlighting the advantages of our probabilistic modeling under uncertainty.

In the synthetic CityCrowd scene, which consists of multiple characters walking in a city environment, we place cameras in a random, CCTV-like configuration to simulate a real-world unconstrained multi-view setup. We report results for both the standard setting and a 50\% spatial sparsity configuration. Consistent with the other datasets, GraphiXS outperforms baselines under these CCTV-style camera distributions.

Overall, we see that different methods are affected differently across datasets, uncertainty types, and uncertainty levels. This is expected as all methods have generality. This is why GraphiXS does not outperform all existing methods across all scenes under all uncertainty settings. However, when analyzing the high-level statistics, GraphiXS provides the best or second-best result in 172 out of a total of 198 individual metrics reported on N3DV dataset in \cref{tab:results_standard_full,tab:results_spatial_sparsity_full,tab:results_sync_temporal_sparsity_full,tab:results_nonsync_temporal_sparsity_full,tab:results_fcam_full} and 159 in 165 on GI dataset in \cref{tab:results_standard_full_gi,tab:results_spatial_sparsity_full_gi,tab:results_sync_temporal_sparsity_full_gi,tab:results_nonsync_temporal_sparsity_full_gi,tab:results_fcam_full_gi}.

\begin{table*}[t!]
\caption{Quantitative comparison on the Neural 3D Video dataset with standard setting.}
\label{tab:results_standard_full}
\centering
\scriptsize
\setlength{\tabcolsep}{2.3pt}
\renewcommand{\arraystretch}{0.95}
\begin{tabularx}{\textwidth}{l|XXXXXX|X|XXXXXX|X|XXXXXX|X}
\toprule
& \multicolumn{7}{c|}{\textbf{PSNR} $\uparrow$} & \multicolumn{7}{c|}{\textbf{DSSIM} $\downarrow$} & \multicolumn{7}{c}{\textbf{LPIPS} $\downarrow$} \\
    & Cof. & Cook & Cut & Flame & Flame & Sear & & Cof. & Cook & Cut & Flame & Flame & Sear & & Cof. & Cook & Cut & Flame & Flame & Sear & \\
    Method & Mart. & Spin. & Beef & Salm. & Steak & Steak & Avg. & Mart. & Spin. & Beef & Salm. & Steak & Steak & Avg. & Mart. & Spin. & Beef & Salm. & Steak & Steak & Avg. \\
\midrule

4DGS-1
& 28.29 & 32.76 & 33.37 & \cellcolor{second}28.86 & 31.48 & 32.29 & 31.18
& \cellcolor{best}0.023 & \cellcolor{second}0.012 & \cellcolor{best}0.010 & \cellcolor{best}0.021 & 0.012 & \cellcolor{second}0.010 & \cellcolor{best}0.015
& 0.075 & 0.042 & 0.041 & 0.070 & 0.036 & 0.035 & 0.051 \\

4DGS-2
& 27.68 & 30.40 & 32.17 & 28.84 & 31.82 & 32.21 & 30.52
& 0.025 & 0.017 & 0.027 & \cellcolor{second}0.022 & 0.012 & 0.012 & 0.019
& 0.084 & 0.055 & 0.040 & 0.081 & 0.043 & 0.042 & 0.060 \\

Ex4DGS
& \cellcolor{best}28.54 & 32.86 & 33.26 & \cellcolor{best}28.95 & 33.17 & \cellcolor{best}33.49 & 31.71
& \cellcolor{best}0.023 & 0.013 & 0.012 & \cellcolor{second}0.022 & \cellcolor{second}0.011 & 0.011 & \cellcolor{best}0.015
& 0.073 & 0.042 & 0.044 & 0.068 & 0.036 & 0.037 & 0.050 \\

FTGS
& 27.36 & 33.28 & \cellcolor{second}33.76 & 28.06 & \cellcolor{second}33.49 & 33.32 & 31.55
& 0.030 & \cellcolor{second}0.012 & \cellcolor{second}0.011 & 0.024 & \cellcolor{best}0.010 & \cellcolor{second}0.010 & \cellcolor{second}0.016
& 0.076 & 0.039 & 0.037 & 0.064 & 0.035 & 0.031 & 0.047 \\
\bottomrule
\end{tabularx}
\end{table*}

\begin{table*}[h]
\caption{Quantitative comparison on the Neural 3D Video dataset with 10\%, 30\%, and 50\% spatial sparsity.}
 \label{tab:results_spatial_sparsity_full}
\centering
\scriptsize
\setlength{\tabcolsep}{2.3pt}
\renewcommand{\arraystretch}{0.95}

\begin{tabularx}{\textwidth}{l|XXXXXX|X|XXXXXX|X|XXXXXX|X}
\toprule
& \multicolumn{7}{c|}{\textbf{PSNR} $\uparrow$} & \multicolumn{7}{c|}{\textbf{DSSIM} $\downarrow$} & \multicolumn{7}{c}{\textbf{LPIPS} $\downarrow$} \\
    & Cof. & Cook & Cut & Flame & Flame & Sear & & Cof. & Cook & Cut & Flame & Flame & Sear & & Cof. & Cook & Cut & Flame & Flame & Sear & \\
    Method & Mart. & Spin. & Beef & Salm. & Steak & Steak & Avg. & Mart. & Spin. & Beef & Salm. & Steak & Steak & Avg. & Mart. & Spin. & Beef & Salm. & Steak & Steak & Avg. \\
\midrule
\multicolumn{21}{c}{\textbf{10\% spatial Sparsity}} \\

4DGS-1
& \cellcolor{second}28.12 & \cellcolor{second}33.25 & \cellcolor{second}33.53 & 28.12 & 32.06 & 33.13 & 31.37
& \cellcolor{second}0.023 & \cellcolor{best}0.011 & \cellcolor{best}0.010 & \cellcolor{second}0.023 & \cellcolor{second}0.011 & \cellcolor{second}0.010 & \cellcolor{best}0.015
& 0.074 & \cellcolor{second}0.040 & 0.039 & 0.068 & 0.036 & 0.034 & 0.049 \\

4DGS-2
& \cellcolor{best}28.73 & 32.14 & 31.90 & \cellcolor{best}28.91 & 31.51 & 32.84 & 31.00
& \cellcolor{best}0.022 & 0.014 & 0.014 & \cellcolor{best}0.021 & 0.012 & 0.011 & \cellcolor{second}0.016
& 0.080 & 0.051 & 0.053 & 0.077 & 0.044 & 0.040 & 0.057 \\

Ex4DGS
& 27.47 & 32.76 & 31.36 & 28.27 & 32.88 & 32.89 & 30.94
& 0.027 & 0.013 & 0.013 & 0.024 & \cellcolor{best}0.010 & 0.011 & \cellcolor{second}0.016
& 0.074 & 0.043 & 0.043 & 0.072 & 0.036 & 0.038 & 0.051 \\

FTGS
& 27.18 & 33.09 & 33.50 & 28.37 & 33.06 & 33.09 & 31.38
& 0.030 & \cellcolor{second}0.012 & \cellcolor{second}0.011 & 0.025 & \cellcolor{second}0.011 & \cellcolor{second}0.010 & \cellcolor{second}0.016
& 0.073 & 0.041 & 0.038 & 0.066 & 0.035 & \cellcolor{second}0.030 & 0.047 \\
\bottomrule
\end{tabularx}
\end{table*}

\begin{table*}[h]
\caption{Quantitative comparison on the Neural 3D Video dataset with temporal sparsity where the 30 FPS videos are trained with 20 FPS and 10 FPS.}
\label{tab:results_sync_temporal_sparsity_full}
\centering
\scriptsize
\setlength{\tabcolsep}{2.3pt}
\renewcommand{\arraystretch}{0.95}

\begin{tabularx}{\textwidth}{l|XXXXXX|X|XXXXXX|X|XXXXXX|X}
\toprule
& \multicolumn{7}{c|}{\textbf{PSNR} $\uparrow$} & \multicolumn{7}{c|}{\textbf{DSSIM} $\downarrow$} & \multicolumn{7}{c}{\textbf{LPIPS} $\downarrow$} \\
    & Cof. & Cook & Cut & Flame & Flame & Sear & & Cof. & Cook & Cut & Flame & Flame & Sear & & Cof. & Cook & Cut & Flame & Flame & Sear & \\
    Method & Mart. & Spin. & Beef & Salm. & Steak & Steak & Avg. & Mart. & Spin. & Beef & Salm. & Steak & Steak & Avg. & Mart. & Spin. & Beef & Salm. & Steak & Steak & Avg. \\
\midrule
\multicolumn{21}{c}{\textbf{20 FPS}} \\

4DGS-1
& 28.19 & 32.29 & 33.35 & 28.48 & 32.18 & 33.05 & 31.26
& \cellcolor{second}0.023 & \cellcolor{second}0.012 & \cellcolor{best}0.010 & \cellcolor{second}0.023 & \cellcolor{best}0.010 & \cellcolor{second}0.010 & \cellcolor{best}0.015
& 0.074 & 0.042 & 0.041 & 0.069 & 0.035 & 0.034 & 0.049 \\

4DGS-2
& \cellcolor{best}28.85 & 31.12 & 31.46 & 28.61 & 31.71 & 31.02 & 30.46
& \cellcolor{best}0.021 & 0.015 & 0.014 & \cellcolor{best}0.022 & 0.012 & 0.012 & \cellcolor{second}0.016
& 0.082 & 0.053 & 0.053 & 0.078 & 0.042 & 0.042 & 0.058 \\

Ex4DGS
& \cellcolor{second}28.27 & 32.46 & 32.84 & \cellcolor{best}28.66 & 32.98 & 33.12 & 31.39
& \cellcolor{second}0.023 & 0.014 & 0.012 & \cellcolor{second}0.023 & \cellcolor{best}0.010 & 0.011 & \cellcolor{best}0.015
& \cellcolor{second}0.072 & 0.044 & 0.044 & 0.069 & 0.035 & 0.037 & 0.050 \\

FTGS
& 27.11 & \cellcolor{second}33.11 & 33.68 & 27.93 & 32.95 & \cellcolor{second}33.57 & 31.39
& 0.031 & \cellcolor{second}0.012 & \cellcolor{second}0.011 & 0.026 & \cellcolor{second}0.011 & \cellcolor{second}0.010 & 0.017
& 0.076 & \cellcolor{second}0.038 & \cellcolor{second}0.037 & 0.065 & \cellcolor{second}0.033 & \cellcolor{second}0.030 & \cellcolor{second}0.046 \\
\bottomrule
\end{tabularx}
\end{table*}

\begin{table*}[h]
\caption{Quantitative comparison on the Neural 3D Video dataset with non-synchronous temporal sparsity where 10\% and 50\% of the views are observed at 20 FPS.}
\label{tab:results_nonsync_temporal_sparsity_full}
\centering
\scriptsize
\setlength{\tabcolsep}{2.3pt}
\renewcommand{\arraystretch}{0.95}

\begin{tabularx}{\textwidth}{l|XXXXXX|X|XXXXXX|X|XXXXXX|X}
\toprule
 & \multicolumn{7}{c|}{\textbf{PSNR} $\uparrow$} & \multicolumn{7}{c|}{\textbf{DSSIM} $\downarrow$} & \multicolumn{7}{c}{\textbf{LPIPS} $\downarrow$} \\
    & Cof. & Cook & Cut & Flame & Flame & Sear & & Cof. & Cook & Cut & Flame & Flame & Sear & & Cof. & Cook & Cut & Flame & Flame & Sear & \\
    Method & Mart. & Spin. & Beef & Salm. & Steak & Steak & Avg. & Mart. & Spin. & Beef & Salm. & Steak & Steak & Avg. & Mart. & Spin. & Beef & Salm. & Steak & Steak & Avg. \\
\midrule
\multicolumn{21}{c}{\textbf{10\% of the views observed at 20 FPS}} \\

4DGS-1
& \cellcolor{second}28.11 & 33.14 & 33.18 & 28.26 & 33.44 & 33.30 & 31.57
& \cellcolor{second}0.023 & \cellcolor{best}0.011 & \cellcolor{second}0.011 & 0.023 & \cellcolor{best}0.010 & \cellcolor{second}0.010 & \cellcolor{second}0.015
& 0.074 & 0.040 & 0.040 & 0.070 & 0.034 & 0.034 & 0.049 \\

4DGS-2
& \cellcolor{best}28.68 & 31.85 & 30.58 & \cellcolor{second}28.80 & 32.49 & 31.70 & 30.68
& \cellcolor{best}0.022 & 0.013 & 0.014 & \cellcolor{second}0.022 & \cellcolor{second}0.012 & 0.012 & 0.016
& 0.082 & 0.051 & 0.054 & 0.082 & 0.043 & 0.043 & 0.059 \\

Ex4DGS
& 27.76 & 32.59 & 32.75 & 28.66 & \cellcolor{best}33.67 & 32.63 & 31.34
& 0.026 & 0.013 & 0.012 & \cellcolor{second}0.022 & \cellcolor{best}0.010 & 0.011 & 0.016
& 0.077 & 0.043 & 0.041 & 0.069 & 0.034 & 0.038 & 0.050 \\

FTGS
& 27.06 & 33.22 & \cellcolor{second}33.82 & 28.24 & 33.17 & 33.15 & 31.44
& 0.031 & \cellcolor{second}0.012 & \cellcolor{second}0.011 & 0.025 & \cellcolor{best}0.010 & \cellcolor{second}0.010 & 0.017
& 0.075 & 0.039 & 0.037 & 0.066 & 0.032 & 0.031 & 0.047 \\
\bottomrule
\end{tabularx}
\end{table*}

\begin{table*}[h]
\caption{Quantitative comparison on the Neural 3D Video dataset with random camera failures.}
\label{tab:results_fcam_full}
\centering
\scriptsize
\setlength{\tabcolsep}{2.3pt}
\renewcommand{\arraystretch}{0.95}

\begin{tabularx}{\textwidth}{l|XXXXXX|X|XXXXXX|X|XXXXXX|X}
\toprule
& \multicolumn{7}{c|}{\textbf{PSNR} $\uparrow$} & \multicolumn{7}{c|}{\textbf{DSSIM} $\downarrow$} & \multicolumn{7}{c}{\textbf{LPIPS} $\downarrow$} \\
    & Cof. & Cook & Cut & Flame & Flame & Sear & & Cof. & Cook & Cut & Flame & Flame & Sear & & Cof. & Cook & Cut & Flame & Flame & Sear & \\
    Method & Mart. & Spin. & Beef & Salm. & Steak & Steak & Avg. & Mart. & Spin. & Beef & Salm. & Steak & Steak & Avg. & Mart. & Spin. & Beef & Salm. & Steak & Steak & Avg. \\
\midrule
\multicolumn{21}{c}{\textbf{Faulty Cam 1}} \\

4DGS-1
& \cellcolor{second}26.58 & 32.31 & 33.16 & \cellcolor{best}27.43 & 32.67 & 32.50 & 30.77
& \cellcolor{second}0.028 & \cellcolor{best}0.013 & \cellcolor{best}0.012 & \cellcolor{best}0.026 & \cellcolor{best}0.011 & \cellcolor{best}0.010 & \cellcolor{best}0.017
& 0.080 & 0.042 & \cellcolor{second}0.040 & 0.077 & 0.034 & 0.034 & 0.051 \\

4DGS-2
& \cellcolor{best}27.28 & 31.77 & 31.84 & \cellcolor{second}27.32 & 32.80 & 32.34 & 30.56
& \cellcolor{best}0.027 & 0.015 & 0.015 & \cellcolor{second}0.027 & 0.013 & 0.012 & \cellcolor{second}0.018
& 0.084 & 0.053 & 0.055 & 0.083 & 0.045 & 0.043 & 0.061 \\

Ex4DGS
& \cellcolor{best}27.28 & 31.96 & 31.87 & \cellcolor{second}27.32 & 32.01 & 32.68 & 30.52
& 0.030 & 0.015 & 0.015 & 0.028 & \cellcolor{second}0.012 & 0.012 & 0.019
& \cellcolor{best}0.077 & 0.047 & 0.046 & 0.074 & 0.038 & 0.039 & 0.053 \\

FTGS
& 25.30 & \cellcolor{second}32.51 & 33.04 & 26.65 & 32.85 & 32.39 & 30.46
& 0.039 & \cellcolor{second}0.014 & \cellcolor{second}0.013 & 0.031 & \cellcolor{second}0.012 & \cellcolor{second}0.011 & 0.020
& 0.083 & \cellcolor{second}0.040 & 0.044 & 0.073 & \cellcolor{second}0.033 & \cellcolor{second}0.032 & 0.051 \\
\bottomrule
\end{tabularx}
\end{table*}

\begin{table*}[t!]
\caption{Quantitative comparison on the Google Immersive Dataset with standard setting.}
\label{tab:results_standard_full_gi}
\centering
\scriptsize
\setlength{\tabcolsep}{2.3pt}
\renewcommand{\arraystretch}{0.95}
\begin{tabularx}{\textwidth}{l|XXXXX|X|XXXXX|X|XXXXX|X}
\toprule
& \multicolumn{6}{c|}{\textbf{PSNR} $\uparrow$} & \multicolumn{6}{c|}{\textbf{DSSIM} $\downarrow$} & \multicolumn{6}{c}{\textbf{LPIPS} $\downarrow$} \\
Method 
& Welder & Flames & Horse & Goats & Alexa & Avg. 
& Welder & Flames & Horse & Goats & Alexa & Avg. 
& Welder & Flames & Horse & Goats & Alexa & Avg. \\
\midrule

4DGS-1
& 24.24 & 28.58 & 26.68 & 25.01 & 29.31 & 26.76
& 0.042 & 0.015 & 0.020 & 0.057 & \cellcolor{second}0.019 & 0.031
& 0.193 & 0.056 & 0.131 & 0.270 & 0.117 & 0.153 \\

4DGS-2
& 22.47 & 27.74 & 23.32 & 20.52 & 25.90 & 23.99
& 0.060 & 0.029 & 0.054 & 0.188 & 0.066 & 0.079
& 0.252 & 0.145 & 0.288 & 0.503 & 0.266 & 0.291 \\

Ex4DGS
& 19.80 & 20.59 & 23.24 & 13.52 & 24.81 & 20.39
& 0.060 & 0.089 & 0.038 & 0.313 & 0.036 & 0.107
& 0.145 & 0.199 & 0.088 & 0.441 & 0.082 & 0.191 \\

FTGS
& \cellcolor{best} 27.61 & 30.18 & 25.22 & 24.63 & \cellcolor{second} 32.40 & 28.01
& \cellcolor{best} 0.015 & 0.010 & 0.013 & 0.066 & \cellcolor{best} 0.008 & 0.022
& 0.061 & 0.032 & 0.059 & 0.195 & 0.021 & 0.074 \\

\midrule
\textbf{Ours (GraphiGS)}
& 25.03 & \cellcolor{second} 31.29 & \cellcolor{best} 29.74 & \cellcolor{second} 25.31 & \cellcolor{best} 32.85 & \cellcolor{second} 28.84
& 0.018 & \cellcolor{second} 0.009 & \cellcolor{second} 0.009 & \cellcolor{second} 0.045 & \cellcolor{best} 0.008 & \cellcolor{second} 0.018
& \cellcolor{best} 0.059 & \cellcolor{second} 0.026 & \cellcolor{second} 0.039 & \cellcolor{second} 0.183 & \cellcolor{best} 0.017 & \cellcolor{second} 0.065 \\
\bottomrule
\end{tabularx}
\end{table*}

\begin{table*}[h]
\caption{Quantitative comparison on the Google Immersive Dataset with 10\%, 30\%, and 50\% spatial sparsity.}
\label{tab:results_spatial_sparsity_full_gi}
\centering
\scriptsize
\setlength{\tabcolsep}{2.3pt}
\renewcommand{\arraystretch}{0.95}

\begin{tabularx}{\textwidth}{l|XXXXX|X|XXXXX|X|XXXXX|X}
\toprule
& \multicolumn{6}{c|}{\textbf{PSNR} $\uparrow$} & \multicolumn{6}{c|}{\textbf{DSSIM} $\downarrow$} & \multicolumn{6}{c}{\textbf{LPIPS} $\downarrow$} \\
Method 
& Welder & Flames & Horse & Goats & Alexa & Avg. 
& Welder & Flames & Horse & Goats & Alexa & Avg. 
& Welder & Flames & Horse & Goats & Alexa & Avg. \\
\midrule

\multicolumn{19}{c}{\textbf{10\% spatial Sparsity}} \\

4DGS-1
& 24.08 & 29.57 & 26.60 & 23.78 & 29.42 & 26.69
& 0.040 & 0.013 & 0.020 & 0.064 & \cellcolor{second}0.018 & 0.031
& 0.183 & 0.049 & 0.127 & 0.276 & 0.105 & 0.148 \\

4DGS-2
& 23.69 & 27.04 & 20.68 & 18.25 & 29.14 & 23.76
& 0.069 & 0.031 & 0.081 & 0.251 & 0.024 & 0.091
& 0.324 & 0.160 & 0.321 & 0.559 & 0.127 & 0.298 \\

Ex4DGS
& 23.20 & 21.10 & 25.12 & 13.35 & 28.24 & 22.20
& 0.038 & 0.070 & 0.022 & 0.317 & 0.020 & 0.093
& 0.104 & 0.168 & 0.062 & 0.466 & 0.038 & 0.168 \\

FTGS
& \cellcolor{best} 27.45 & \cellcolor{second} 30.08 & 26.73 & 22.54 & \cellcolor{second} 32.58 & 27.88
& \cellcolor{best} 0.016 & \cellcolor{second}0.011 & \cellcolor{second}0.017 & 0.104 & \cellcolor{best} 0.008 & 0.031
& \cellcolor{best} 0.061 & \cellcolor{second}0.030 & 0.102 & \cellcolor{second}0.257 & \cellcolor{second} 0.018 & 0.094 \\

\midrule

\textbf{Ours (GraphiGS)}
& 25.81 & 29.85 & \cellcolor{best} 29.69 & \cellcolor{second} 27.52 & 31.54 & \cellcolor{second} 28.88
& 0.020 & \cellcolor{second}0.011 & \cellcolor{best} 0.009 & \cellcolor{second} 0.031 & \cellcolor{best}0.008 & \cellcolor{second} 0.016
& \cellcolor{second}0.077 & \cellcolor{best} 0.028 & \cellcolor{best} 0.036 & \cellcolor{best} 0.128 & \cellcolor{second}0.018 & \cellcolor{second} 0.057 \\
\bottomrule
\end{tabularx}
\end{table*}

\begin{table*}[h]
\caption{Quantitative comparison on the Google Immersive Dataset with temporal sparsity where the 30 FPS videos are trained with 20 FPS and 10 FPS.}
\label{tab:results_sync_temporal_sparsity_full_gi}
\centering
\scriptsize
\setlength{\tabcolsep}{2.3pt}
\renewcommand{\arraystretch}{0.95}

\begin{tabularx}{\textwidth}{l|XXXXX|X|XXXXX|X|XXXXX|X}
\toprule
& \multicolumn{6}{c|}{\textbf{PSNR} $\uparrow$} & \multicolumn{6}{c|}{\textbf{DSSIM} $\downarrow$} & \multicolumn{6}{c}{\textbf{LPIPS} $\downarrow$} \\
Method 
& Welder & Flames & Horse & Goats & Alexa & Avg. 
& Welder & Flames & Horse & Goats & Alexa & Avg. 
& Welder & Flames & Horse & Goats & Alexa & Avg. \\
\midrule

\multicolumn{19}{c}{\textbf{20 FPS}} \\

4DGS-1
& 22.90 & 29.13 & 25.85 & 23.15 & 29.22 & 26.05
& 0.053 & 0.018 & 0.027 & 0.074 & 0.025 & 0.039
& 0.206 & 0.064 & 0.144 & 0.299 & 0.126 & 0.167 \\

4DGS-2
& 23.60 & 25.50 & 19.18 & 20.29 & 27.29 & 23.17
& 0.070 & 0.043 & 0.076 & 0.168 & 0.036 & 0.078
& 0.317 & 0.215 & 0.314 & 0.399 & 0.177 & 0.284 \\

Ex4DGS
& 21.50 & 24.12 & 25.83 & 13.15 & 20.60 & 21.04
& 0.056 & 0.048 & 0.020 & 0.295 & 0.053 & 0.095
& 0.132 & 0.124 & 0.056 & 0.467 & 0.101 & 0.176 \\

FTGS
& 25.80 & \cellcolor{best} 30.35 & 27.22 & \cellcolor{best} 24.63 & 30.94 & 27.79
& 0.024 & \cellcolor{best} 0.010 & 0.014 & \cellcolor{second} 0.065 & \cellcolor{second} 0.009 & 0.025
& 0.100 & \cellcolor{second} 0.030 & 0.078 & 0.204 & \cellcolor{second}0.022 & \cellcolor{second}0.087 \\

\midrule

\textbf{Ours (Gaussian)}
& \cellcolor{best} 27.30 & \cellcolor{second} 30.22 & \cellcolor{best} 29.10 & \cellcolor{second} 24.55 & \cellcolor{best} 32.21 & \cellcolor{best} 28.68
& \cellcolor{best} 0.018 & \cellcolor{second}0.011 & \cellcolor{best} 0.010 & 0.072 & \cellcolor{best} 0.008 & \cellcolor{second} 0.024
& \cellcolor{best} 0.063 & 0.035 & \cellcolor{best} 0.044 & \cellcolor{second} 0.178 & \cellcolor{best} 0.020 & \cellcolor{best} 0.068 \\
\bottomrule
\end{tabularx}
\end{table*}

\begin{table*}[h]
\caption{Quantitative comparison on the Google Immersive Dataset with non-synchronous temporal sparsity where 10\% and 50\% of the views are observed at 20 FPS.}
\label{tab:results_nonsync_temporal_sparsity_full_gi}
\centering
\scriptsize
\setlength{\tabcolsep}{2.3pt}
\renewcommand{\arraystretch}{0.95}

\begin{tabularx}{\textwidth}{l|XXXXX|X|XXXXX|X|XXXXX|X}
\toprule
 & \multicolumn{6}{c|}{\textbf{PSNR} $\uparrow$} & \multicolumn{6}{c|}{\textbf{DSSIM} $\downarrow$} & \multicolumn{6}{c}{\textbf{LPIPS} $\downarrow$} \\
Method
& Welder & Flames & Horse & Goats & Alexa & Avg.
& Welder & Flames & Horse & Goats & Alexa & Avg.
& Welder & Flames & Horse & Goats & Alexa & Avg. \\
\midrule
\multicolumn{19}{c}{\textbf{10\% of the views observed at 20 FPS}} \\

4DGS-1
& 23.60 & 29.36 & 26.37 & 24.60 & 29.35 & 26.66
& 0.043 & \cellcolor{second}0.015 & 0.022 & 0.065 & 0.019 & 0.033
& 0.193 & 0.054 & 0.130 & 0.289 & 0.108 & 0.155 \\

4DGS-2
& 24.01 & 26.23 & 19.28 & 19.88 & 24.04 & 22.69
& 0.057 & 0.049 & 0.075 & 0.165 & 0.098 & 0.089
& 0.271 & 0.230 & 0.297 & 0.419 & 0.207 & 0.285 \\

Ex4DGS
& 24.10 & 23.94 & 21.77 & 13.09 & 27.39 & 22.06
& 0.031 & 0.056 & 0.025 & 0.333 & 0.025 & 0.094
& 0.099 & 0.128 & 0.070 & 0.502 & 0.062 & 0.172 \\

FTGS
& \cellcolor{second} 25.38 & \cellcolor{best} 31.36 & 26.51 & 22.16 & \cellcolor{second} 32.04 & 27.49
& \cellcolor{second} 0.020 & \cellcolor{best} 0.009 & 0.013 & 0.094 & \cellcolor{second}0.009 & 0.029
& \cellcolor{second} 0.064 & \cellcolor{best} 0.026 & 0.059 & 0.223 & 0.023 & 0.079 \\

\midrule
\textbf{Ours (GraphiGS)}
& \cellcolor{best} 27.30 & 30.89 & \cellcolor{second} 29.48 & \cellcolor{second} 26.01 & 31.60 & \cellcolor{best} 29.06
& \cellcolor{best} 0.015 & \cellcolor{best} 0.009 & \cellcolor{best} 0.008 & \cellcolor{best} 0.033 & \cellcolor{best} 0.008 & \cellcolor{best} 0.015
& \cellcolor{best} 0.056 & \cellcolor{second}0.029 & \cellcolor{best} 0.037 & \cellcolor{best} 0.137 & \cellcolor{best} 0.019 & \cellcolor{best} 0.055 \\
\bottomrule
\end{tabularx}
\end{table*}

\begin{table*}[h]
\caption{Quantitative comparison on the Google Immersive Dataset with random camera failures.}
\label{tab:results_fcam_full_gi}
\centering
\scriptsize
\setlength{\tabcolsep}{2.3pt}
\renewcommand{\arraystretch}{0.95}

\begin{tabularx}{\textwidth}{l|XXXXX|X|XXXXX|X|XXXXX|X}
\toprule
& \multicolumn{6}{c|}{\textbf{PSNR} $\uparrow$} & \multicolumn{6}{c|}{\textbf{DSSIM} $\downarrow$} & \multicolumn{6}{c}{\textbf{LPIPS} $\downarrow$} \\
Method 
& Welder & Flames & Horse & Goats & Alexa & Avg.
& Welder & Flames & Horse & Goats & Alexa & Avg.
& Welder & Flames & Horse & Goats & Alexa & Avg. \\
\midrule

\multicolumn{19}{c}{\textbf{Faulty Cam 1}} \\

4DGS-1
& 23.26 & 27.91 & 25.87 & 23.37 & 30.05 & 26.09
& 0.046 & 0.020 & 0.023 & 0.070 & \cellcolor{second}0.018 & 0.035
& 0.198 & 0.060 & 0.137 & 0.286 & 0.110 & 0.158 \\

4DGS-2
& 19.54 & 23.60 & 18.93 & 17.50 & 29.17 & 21.75
& 0.142 & 0.062 & 0.099 & 0.206 & 0.023 & 0.106
& 0.455 & 0.245 & 0.327 & 0.452 & 0.103 & 0.316 \\

Ex4DGS
& 21.31 & 22.12 & 20.64 & 13.56 & 26.44 & 20.81
& 0.051 & 0.075 & 0.053 & 0.302 & 0.032 & 0.103
& 0.125 & 0.178 & 0.113 & 0.446 & 0.068 & 0.186 \\

FTGS
& \cellcolor{best} 26.33 & \cellcolor{second}30.75 & \cellcolor{second} 28.74 & \cellcolor{second}23.90 & \cellcolor{best} 31.55 & \cellcolor{second} 28.25
& \cellcolor{best} 0.019 & \cellcolor{second} 0.010 & \cellcolor{second}0.011 & \cellcolor{second}0.075 & \cellcolor{best}0.009 & \cellcolor{second} 0.025
& \cellcolor{second} 0.069 & 0.040 & 0.047 & \cellcolor{second}0.203 & \cellcolor{best} 0.021 & 0.076 \\

\midrule
\textbf{Ours (GraphiGS)}
& \cellcolor{second}25.64 & \cellcolor{best} 31.10 & \cellcolor{best} 28.79 & \cellcolor{best} 26.22 & 30.68 & \cellcolor{best} 28.49
&\cellcolor{second} 0.020 & \cellcolor{best} 0.009 & \cellcolor{best} 0.010 & \cellcolor{best} 0.034 & \cellcolor{best} 0.009 & \cellcolor{best} 0.016
& \cellcolor{best} 0.063 & \cellcolor{best} 0.026 & \cellcolor{best} 0.038 & \cellcolor{best} 0.133 & \cellcolor{best} 0.021 & \cellcolor{best} 0.056 \\
\bottomrule
\end{tabularx}
\end{table*}

\begin{table}[t!]
\caption{Quantitative comparison on the CityCrowd dataset (scene: city\_01) under the Standard and 50\% Spatial Sparsity settings.}
\label{tab:results_citycrowd}
\centering
\scriptsize
\setlength{\tabcolsep}{2.8pt}
\renewcommand{\arraystretch}{1.05}

\begin{tabular}{l|ccc|ccc}
\toprule
\multirow{2}{*}[-1.0ex]{\textbf{Method}}
& \multicolumn{3}{c}{\textbf{Standard}}
& \multicolumn{3}{c}{\textbf{50\% Spatial Sparsity}} \\
\cmidrule(lr){2-4} \cmidrule(lr){5-7}
& \textbf{PSNR}$\uparrow$
& \textbf{DSSIM}$\downarrow$
& \textbf{LPIPS}$\downarrow$
& \textbf{PSNR}$\uparrow$
& \textbf{DSSIM}$\downarrow$
& \textbf{LPIPS}$\downarrow$ \\
\midrule
4DGS-1
& 27.36 & 0.028 & 0.100
& 24.33 & 0.040 & 0.112 \\

4DGS-2
& 15.71 & 0.276 & 0.512
& 15.17 & 0.275 & 0.504 \\

Ex4DGS
& 19.34 & 0.149 & 0.301
& 17.36 & 0.243 & 0.374 \\

FTGS
& \cellcolor{second}28.09 & 0.019 & 0.068
& 25.96 & 0.029 & \cellcolor{second}0.083 \\

\midrule
\textbf{Ours (GraphiGS)}
& 28.02 & \cellcolor{second}0.017 & \cellcolor{second}0.061
& \cellcolor{second}25.97 & \cellcolor{second}0.029 & \cellcolor{best}0.076 \\

\textbf{Ours (GraphiTS)}
& \cellcolor{best}28.17 & \cellcolor{best}0.016 & \cellcolor{best}0.057
& \cellcolor{best}26.42 & \cellcolor{best}0.028 & \cellcolor{best}0.076 \\
\bottomrule
\end{tabular}
\end{table}

\subsection{Qualitative Results} 
We present further visual comparisons on the N3DV dataset in \cref{fig:qualitative_page_0,fig:qualitative_page_1}, on the GI dataset in \cref{fig:qualitative_page_2,fig:qualitative_page_3}, and on the CityCrowd dataset in \cref{fig:qualitative_page_4}. Consistent with the findings in the main manuscript, in general, GraphiXS recovers finer high-frequency details across all uncertainty settings. Notably, dynamic scene components are reconstructed with significantly higher fidelity compared to baselines. This is particularly evident as temporal sparsity increases. The crops highlighting the subject’s hands and motion in the rows corresponding to 10 FPS, 10\% @ 20 FPS, 50\% @ 20 FPS, and Random Camera Failures 1 in \cref{fig:qualitative_page_0,fig:qualitative_page_1}. We attribute this performance gap to the expressiveness of our higher-order motion model. 

A trade-off of our approach is that when most cameras are directed away from the primary region of interest, the model assigns lower confidence to under-observed areas, which can lead to reduced quality in those regions. This effect is visible in the Flame Salmon and Coffee Martini scenes in \cref{fig:qualitative_page_0,fig:qualitative_page_1}, where outdoor regions have lower visibility, while the representation budget is concentrated on the well-observed areas, resulting in higher-quality reconstruction and motion fidelity in those regions.

\subsection{Higher Order Motion} 
\begin{figure}[h!]
    \centering
    \includegraphics[width=0.90\linewidth]{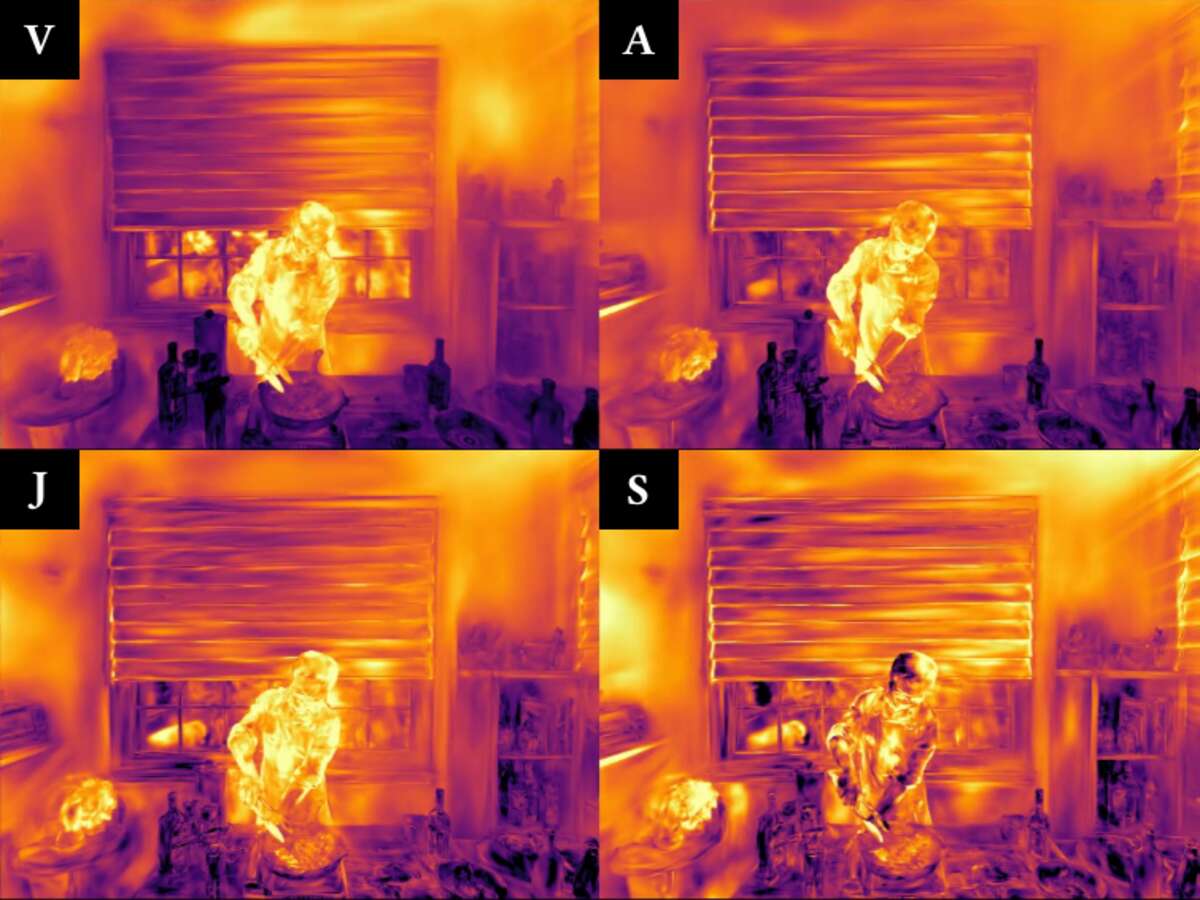}
    \caption{Magnitude of motion derivatives. Each image is independently normalized. V, A, J, and S denote the magnitudes of velocity, acceleration, jerk, and snap, respectively.}
    \label{fig:motion_model_visualization}
\end{figure}

\Cref{fig:motion_model_visualization} visualizes the magnitudes of the motion derivatives, providing empirical evidence for our choice of modeling the component dynamics up to the fourth order. As expected, dynamic regions exhibit strong responses in the lower-order terms, corresponding to velocity and acceleration. As the derivative order increases, the magnitudes progressively decay. By the fourth order (snap), the response in moving regions is substantially attenuated, appearing as darker areas in the visualization. This systematic decay suggests that the dominant motion dynamics in the data are well captured by the first four derivatives. Consequently, extending the model beyond the fourth order is unlikely to yield meaningful improvements in reconstruction quality, while incurring additional computational and storage costs.


\subsection{Training and Storage Requirements}
We report the training time, storage requirements, training memory usage, and inference FPS of GraphiXS, and compare them against baseline methods in \cref{tab:performance}. All training times are measured on an NVIDIA RTX 4090 GPU. GraphiXS has slightly higher training cost primarily due to the component confidence computation, which requires multi-view rendering of the scene at each iteration. Overall, this is expected as the introduction of stochasticity into various steps of 4DGS requires more computation for the optimization. We argue the extra time cost is meaningful to enable the model to accommodate different types of data uncertainty. For inference, GraphiXS achieves a mid-range FPS among the baselines, primarily due to the increased complexity of its motion modeling.

In terms of storage, GraphiXS exhibits a moderate footprint. Compared to FTGS, the increased storage consumption is mainly caused by the additional parameters introduced by the higher-order motion model and the per-component confidence scores. Among the baselines, 4DGS-2 is the most storage-efficient, as it maintains only a canonical set of components and employs an MLP-based deformation model to generate time-dependent components. However, this storage efficiency comes at the cost of reduced visual quality compared to the other methods. At the opposite end of the spectrum, 4DGS-1 represents scene motion using a large number of very short-lived Gaussians. These components are limited to zeroth-order dynamics, resulting in minimal temporal sharing compared to Ex4DGS, FTGS, and GraphiXS. GraphiXS occupies the middle ground in this trade-off space, leveraging short-lived yet temporally shared components to balance storage efficiency and reconstruction quality.

\begin{table}[tb]
\caption{Average performance comparison on N3DV dataset \cite{li2022neural} using standard setting.}
\label{tab:performance}
\centering
\scriptsize
\setlength{\tabcolsep}{2.8pt}
\renewcommand{\arraystretch}{1.05}

\begin{tabular}{l|cccc}
\toprule
\multirow{1}{*}{\textbf{Method}}
& \begin{tabular}[c]{@{}c@{}}\textbf{Storage Size}$\downarrow$\end{tabular}
& \begin{tabular}[c]{@{}c@{}}\textbf{Training Time}$\downarrow$\end{tabular}
& \begin{tabular}[c]{@{}c@{}}\textbf{Training Memory}$\downarrow$\end{tabular}
& \begin{tabular}[c]{@{}c@{}}\textbf{Inference FPS}$\uparrow$\end{tabular} \\
\midrule

4DGS-1
& 3128 MB & 1.5 Hours & 25814 MB & 120 FPS \\

4DGS-2
& 90 MB & 40 Minutes & 15428 MB & 34 FPS \\

Ex4DGS
& 115 MB & 1 Hour & 3759 MB & 46 FPS \\

FTGS
& 366 MB & 2.5 Hours & 3751 MB & 126 FPS \\

\midrule

\textbf{Ours (GraphiGS)}
& 423 MB & 4.5 Hours & 4701 MB & 113 FPS \\

\textbf{Ours (GraphiTS)}
& 429 MB & 4.5 Hours & 6471 MB & 89 FPS \\

\bottomrule
\end{tabular}
\end{table}

\section{Additional Details of Methodology}
This section provides additional details about GraphiXS and hyperparameters used in the trainings.

\subsection{Variants of GraphiXS}
\label{sec:graphixs_variants}

As established in our generative formulation, GraphiXS is a generalized framework in which the component $\delta_{\theta}$ is modeled as a probabilistic distribution parameterized by $\theta$, rather than being restricted to a fixed geometric primitive. This design enables flexible instantiations of the framework using different distribution families.

\paragraph{GraphiGS}
As the first variant of GraphiXS we initialize $\delta_{\theta}$ using 3D Gaussian primitives and refer this version as GraphiGS. In this variant, the component $\delta$ is defined as a Gaussian distribution parameterized by a mean $\boldsymbol{\mu} \in \mathbb{R}^3$ and a covariance matrix $\boldsymbol{\Sigma} \in \mathbb{R}^{3 \times 3}$. This formulation is consistent with the standard 3D Gaussian Splatting paradigm, where density decays exponentially with distance from the mean. Formally, the distribution for a single component in GraphiGS is given by
\begin{equation}
\delta_{\mathrm{GS}}(\mathbf{x}; \boldsymbol{\mu}, \boldsymbol{\Sigma})
=\exp\!\left(-\frac{1}{2}(\mathbf{x} - \boldsymbol{\mu})^{\top}\boldsymbol{\Sigma}^{-1} (\mathbf{x} - \boldsymbol{\mu})\right).
\end{equation}

\paragraph{GraphiTS}
While Gaussian primitives provide compact spatial support, they may be limited in capturing broader uncertainty regions. To address this, we introduce a second variant, GraphiTS, which instantiates $\delta$ using Student's t distribution. Following the formulation in 3DSSS \cite{zhu20253d}, we augment the parameterization with a learnable degrees-of-freedom parameter $\nu \in [1, +\infty)$, in addition to $\boldsymbol{\mu}$ and $\boldsymbol{\Sigma}$. The parameter $\nu$ explicitly controls the tail heaviness of the distribution, allowing GraphiTS to smoothly interpolate between heavy-tailed Cauchy-like behavior and Gaussian-like behavior. The corresponding distribution is defined as
\begin{equation}
\delta_{\mathrm{TS}}(\mathbf{x}; \boldsymbol{\mu}, \boldsymbol{\Sigma}, \nu)
=\left[1 + \frac{1}{\nu}(\mathbf{x} - \boldsymbol{\mu})^{\top}\boldsymbol{\Sigma}^{-1}(\mathbf{x} - \boldsymbol{\mu}) \right]^{-\frac{\nu + 3}{2}}.
\end{equation}

\paragraph{More Variants}
Beyond the two variants described above, the probabilistic formulation of GraphiXS naturally supports alternative distribution families for the component $\delta$, as long as they can be parameterized by the shared latent variables $\theta$. This flexibility enables the incorporation of distributions with different support characteristics and uncertainty profiles. Beta distribution based primitives, as introduced in \cite{liu2025universal, liu2025deformable}, may serve as a promising candidate for a future GraphiBS variant.

\subsection{Prior on Color}
To capture the view-dependent appearance of the moving components, we condition the color $\PrimitiveColor$ on $\mean(t)$ by \cite{wang2025freetimegs}:
\begin{equation}
\begin{aligned}
\PrimitiveColor &= \sum_{l=0}^{L} \sum_{m=-l}^{l} sh_{lm}\, Y_{lm}\!\bigl(D(\mu(t))\bigr) 
\label{eq:color}
\end{aligned}
\end{equation}
where $\PrimitiveColor$ is the time and view dependent color, $L$ is the degree of the spherical harmonics, $D$ is the viewing direction at position $\mean(\CurrentTime)$, and $Y$ is the spherical harmonics basis function evaluated at the viewing direction $D(\mean(\CurrentTime))$. By explicitly conditioning the color evaluation on the dynamic position $\mu(t)$, we ensure that the view-dependent effects remain consistent with the component's learned trajectory and higher-order dynamics.
 
\subsection{Sampling via Stochastic Gradient Hamiltonian Monte Carlo}
To optimize the posterior distribution defined in the main manuscript, we adopt Stochastic Gradient Hamiltonian Monte Carlo (SGHMC) \cite{zhu20253d}. SGHMC is applied jointly to all learnable component parameters $\theta$, enabling efficient exploration of the posterior. This choice is motivated by the highly non-convex solution space induced by our objective, where parameters governing shape, motion, and appearance are tightly coupled.

\subsection{Adding and Recycling Components in 4D}
During optimization, components with near-zero opacity contribute negligibly to the rendered scene while unnecessarily consuming the component budget. We identify such components using an opacity threshold and mark them as unused. Unused components are recycled regardless of budget, while new components are added only when the active set falls below the budget. We regulate both component addition and recycling using a scalar importance score
\begin{equation}
\score_\components^t = \lambda_{sg},\bar{g}_\components^t + \lambda{so}, \opacity_\components^t,
\end{equation}
where $\bar{g}_\components^t$ denotes the accumulated positional gradient of components $\components$ at time $\CurrentTime$, $\opacity_\components^t$ its opacity, and $\lambda{sg}$ and $\lambda_{so}$ are weighting coefficients. Active components are sampled according to the resulting importance distribution

\begin{equation}
\tau_\components^t = \frac{\score_\components^t}{\sum^\mat{P^t}_\components \score_\components^t + \epsilon}.
\end{equation}

The sampled components guide both the placement of newly added components and the reassignment of recycled ones. As a result, component capacity is dynamically focused on regions that most strongly influence the 4D reconstruction.

\subsection{Initialization}
Following \cite{wang2025freetimegs}, we initialize our model from temporally coherent 4D point clouds reconstructed from multi-view image sequences. For each video frame, dense 3D point clouds are reconstructed independently by matching keypoints across views using RoMa \cite{edstedt2024roma} and triangulating the resulting correspondences. These per-frame point clouds are subsequently merged into a unified 4D point set by establishing temporal correspondences via k-nearest-neighbor matching between points in successive frames. The temporal coordinate of each point is initialized to the corresponding video timestamp, while point velocities are derived from inter-frame displacements along the temporal correspondences. Higher-order motion derivatives are initialized using a temporal binning scheme. For each derivative order, we compute deviations from the preceding order within fixed temporal bins and assign the resulting statistics to the corresponding points. This procedure yields bounded and physically plausible motion initialization.

\subsection{Hyperparameters}
We implement our framework using PyTorch and CUDA. Optimization of the component parameters $\theta$ is performed using SGHMC with a friction coefficient of $C = 120$ and a noise scale of $1.0$. To stabilize early training, we employ a burn-in phase of $7{,}000$ iterations, during which the friction coefficient is increased to $C_{\text{burn-in}} = 5 \times 10^{5}$ in order to suppress excessive oscillations. For higher-order dynamics, the standard deviation of the Brownian motion noise is set to $\epsilon = 1 \times 10^{-4}$.

The learning rates for dynamics-related parameters—velocity $v$, acceleration $a$, jerk $j$, snap $s$, and component duration $u$—are exponentially annealed from an initial value of $1 \times 10^{-2}$ to a final value of $1 \times 10^{-4}$ over the course of training.

We assign fixed weights to the loss terms and regularization priors. The component confidence regularization weight $\lambda_{\alpha}$ for $\mathcal{L}_{\alpha}$ is set to $1 \times 10^{-3}$. The temporal opacity regularization weight is set to $0.01$, and the scale regularization weight associated with $\epsilon_{\boldsymbol{\Sigma}}$ is also set to $0.01$. For the smoothness prior on the dynamics derivatives, the weights for velocity, acceleration, jerk, and snap are set uniformly to $\lambda_h = 1 \times 10^{-4}$. For importance sampling in component addition and recycling, we assign equal weights to the gradient and opacity terms, with $\lambda_{sg} = 0.5$ and $\lambda_{so} = 0.5$, respectively. Across all GraphiXS experiments, we cap the total number of components at $1.5$ million.

\subsection{GraphiXS as an Upgrade}
GraphiXS is a generalized probabilistic framework that can be applied to a wide range of 4D Gaussian Splatting (4DGS) representations. In contrast, most existing 4DGS methods formulate reconstruction as a deterministic optimization of photometric error:
\begin{equation}
\min \ \mathcal{L}_1 + \lambda \, \mathcal{L}_{\text{D-SSIM}}.
\label{eq:deterinisticLoss}
\end{equation}
To upgrade a baseline 4DGS method using GraphiXS, we reformulate its training objective as a Maximum a Posteriori (MAP) estimation problem, as defined in the main paper. This reformulation involves two main adaptations: reinterpreting the likelihood and incorporating stochastic priors.

\paragraph{General Adaptation Strategy}
First, the baseline rendering loss is converted into an energy-based likelihood distribution. Our energy-based likelihood model in the main paper can be directly used, as it includes \Cref{eq:deterinisticLoss}:
\begin{align}
    &P(\image \mid \beta, \Ray, \alpha, \CameraPose, \Times) \propto exp(-\sum^\NumberOfCams_c\sum^\NumberOfFrames_t L_{img} (I_c^t)) \nonumber \\
    L_{img} &= (1-\epsilon_{D-SSIM}) L_1 + \epsilon_{D-SSIM} L_{D-SSIM} \nonumber \\
    &+ \epsilon_\opacity \sum_i ||\opacity_i||_1 + \epsilon_{\cov}\sum_i\sum_j ||\sqrt{\lambda_{i,j}}||_1
    \label{eq:imageDis}
\end{align}
where $L_1$ and $L_{D-SSIM}$ are the $L_1$ norm and the structural similarity loss between the reconstructed image and the ground-truth.  $\lambda$s are the eigenvalues of $\cov$. The regularization applied to the opacity ensures that the opacity is big only when a component is absolutely needed. The regularization on $\lambda$ ensures the model uses components as spiky as possible (i.e. small variances). Together, the regularization terms minimize the needed number of components.

Second, we introduce the \emph{Component Confidence} term,
\begin{equation}
P(\alpha \mid \delta_\theta, C, T),
\end{equation}
which is incorporated as a regularization term $\mathcal{L}_\alpha$. This term encourages the model to account for data uncertainty by maximizing the joint probability of component visibility across all available views and timestamps per batch. Regardless the formulation and the implementation of the base methods, this component confidence term can be computed. 

Finally, the shape prior $P(\Sigma_t^p)$ can also be applied to ensure consistency of primitive covariances over time. This term is also general as long as the component is a probabilistic distribution, as the covariance matrix is a common parameter.

Methods such as FTGS \cite{wang2025freetimegs}, Ex4DGS \cite{lee2024fully}, and 4DGS-1 \cite{yang2023gs4d} use explicit 3D/4D Gaussians or similar primitives that move dynamically, allowing the above adaptation strategy to be applied directly. As an example in the paper, we upgraded FTGS. By contrast, methods like 4DGS-2 \cite{wu20244d}, which rely on a canonical Gaussian set deformed via a neural field (\eg, HexPlane or MLP), cannot be upgraded directly. In these cases, GraphiXS priors can be applied to the displaced components at each sampled timestamp, with gradients backpropagated through the deformation network.

\subsection{Deterministic Variant of GraphiXS}
In our ablation study, we introduce a deterministic variant of GraphiXS to validate the effectiveness of our framework. In this variant, we replace the SGHMC optimizer with standard Adam, thereby removing Hamiltonian Monte Carlo noise from gradient updates. We also disable the component confidence loss, eliminating the incentive to maintain visibility across multiple viewpoints, and remove Brownian motion noise during rasterization. Furthermore, stochastic multinomial sampling for component addition and recycling during densification is replaced with a deterministic strategy that selects the top-$N$ components based on the same scoring criteria.
\balance

\begin{figure*}[h]
    \centering

    \begin{minipage}{\colw}\centering\scriptsize 4DGS-1\end{minipage}
    \begin{minipage}{\colw}\centering\scriptsize 4DGS-2\end{minipage}
    \begin{minipage}{\colw}\centering\scriptsize Ex4DGS\end{minipage}
    \begin{minipage}{\colw}\centering\scriptsize FTGS\end{minipage}
    \begin{minipage}{\colw}\centering\scriptsize \textbf{Ours (GraphiGS)}\end{minipage}
    \begin{minipage}{\colw}\centering\scriptsize \textbf{Ours (GraphiTS)}\end{minipage}
    \begin{minipage}{\colw}\centering\scriptsize GT\end{minipage}

    \par\vspace{4pt}
    \hrule
    \par\vspace{6pt}

    \figrow{Standard Setting}{figs/full_page/full_page_std}
    \figrow{10\% spatial Sparsity}{figs/full_page/full_page_sparse_90}
    \figrow{30\% spatial Sparsity}{figs/full_page/full_page_sparse_70}
    \figrow{50\% spatial Sparsity}{figs/full_page/full_page_sparse_50}
    \figrow{20 FPS}{figs/full_page/full_page_fps_20}

    \caption{
        Qualitative comparison on N3DV dataset. Each row corresponds to a different experimental setting. Columns show six competing methods and the ground truth (GT). Per-region PSNR values are reported in the top-right corner of each cropped region.
    }
    \label{fig:qualitative_page_0}
\end{figure*}

\begin{figure*}[h!]
    \centering

    \begin{minipage}{\colw}\centering\scriptsize 4DGS-1\end{minipage}
    \begin{minipage}{\colw}\centering\scriptsize 4DGS-2\end{minipage}
    \begin{minipage}{\colw}\centering\scriptsize Ex4DGS\end{minipage}
    \begin{minipage}{\colw}\centering\scriptsize FTGS\end{minipage}
    \begin{minipage}{\colw}\centering\scriptsize \textbf{Ours (GraphiGS)}\end{minipage}
    \begin{minipage}{\colw}\centering\scriptsize \textbf{Ours (GraphiTS)}\end{minipage}
    \begin{minipage}{\colw}\centering\scriptsize GT\end{minipage}

    \par\vspace{4pt}
    \hrule
    \par\vspace{6pt}

    \figrow{10 FPS}{figs/full_page/full_page_fps_10}
    \figrow{10\% of the views observed at 20 FPS}{figs/full_page/full_page_nosync_90}
    \figrow{50\% of the views observed at 20 FPS}{figs/full_page/full_page_nosync_50}
    \figrow{Random Camera Failures 1}{figs/full_page/full_page_fcam}
    \figrow{Random Camera Failures 2}{figs/full_page/full_page_fcam2}
    \figrow{Random Camera Failures 3}{figs/full_page/full_page_fcam3}

    \caption{
        Qualitative comparison on N3DV dataset. Each row corresponds to a different experimental setting. Columns show six competing methods and the ground truth (GT). Per-region PSNR values are reported in the top-right corner of each cropped region.
    }
    \label{fig:qualitative_page_1}
\end{figure*}

\begin{figure*}[h]
    \centering

    \begin{minipage}{\colw}\centering\scriptsize 4DGS-1\end{minipage}
    \begin{minipage}{\colw}\centering\scriptsize 4DGS-2\end{minipage}
    \begin{minipage}{\colw}\centering\scriptsize Ex4DGS\end{minipage}
    \begin{minipage}{\colw}\centering\scriptsize FTGS\end{minipage}
    \begin{minipage}{\colw}\centering\scriptsize \textbf{Ours (GraphiGS)}\end{minipage}
    \begin{minipage}{\colw}\centering\scriptsize \textbf{Ours (GraphiTS)}\end{minipage}
    \begin{minipage}{\colw}\centering\scriptsize GT\end{minipage}

    \par\vspace{4pt}
    \hrule
    \par\vspace{6pt}

    \figrow{Standard Setting}{figs/full_page_gi/full_page_std}
    \figrow{10\% spatial Sparsity}{figs/full_page_gi/full_page_sparse_90}
    \figrow{30\% spatial Sparsity}{figs/full_page_gi/full_page_sparse_70}
    \figrow{50\% spatial Sparsity}{figs/full_page_gi/full_page_sparse_50}
    \figrow{20 FPS}{figs/full_page_gi/full_page_fps_20}

    \caption{
        Qualitative comparison on GI dataset. Each row corresponds to a different experimental setting. Columns show six competing methods and the ground truth (GT). Per-region PSNR values are reported in the top-right corner of each cropped region.
    }
    \label{fig:qualitative_page_2}
\end{figure*}

\begin{figure*}[h!]
    \centering

    \begin{minipage}{\colw}\centering\scriptsize 4DGS-1\end{minipage}
    \begin{minipage}{\colw}\centering\scriptsize 4DGS-2\end{minipage}
    \begin{minipage}{\colw}\centering\scriptsize Ex4DGS\end{minipage}
    \begin{minipage}{\colw}\centering\scriptsize FTGS\end{minipage}
    \begin{minipage}{\colw}\centering\scriptsize \textbf{Ours (GraphiGS)}\end{minipage}
    \begin{minipage}{\colw}\centering\scriptsize \textbf{Ours (GraphiTS)}\end{minipage}
    \begin{minipage}{\colw}\centering\scriptsize GT\end{minipage}

    \par\vspace{4pt}
    \hrule
    \par\vspace{6pt}

    \figrow{10 FPS}{figs/full_page_gi/full_page_fps_10}
    \figrow{10\% of the views observed at 20 FPS}{figs/full_page_gi/full_page_nosync_90}
    \figrow{50\% of the views observed at 20 FPS}{figs/full_page_gi/full_page_nosync_50}
    \figrow{Random Camera Failures 1}{figs/full_page_gi/full_page_fcam}
    \figrow{Random Camera Failures 2}{figs/full_page_gi/full_page_fcam2}
    \figrow{Random Camera Failures 3}{figs/full_page_gi/full_page_fcam3}

    \caption{
        Qualitative comparison on GI dataset. Each row corresponds to a different experimental setting. Columns show six competing methods and the ground truth (GT). Per-region PSNR values are reported in the top-right corner of each cropped region.
    }
    \label{fig:qualitative_page_3}
\end{figure*}

\begin{figure*}[h!]
    \centering

    \begin{minipage}{\colw}\centering\scriptsize 4DGS-1\end{minipage}
    \begin{minipage}{\colw}\centering\scriptsize 4DGS-2\end{minipage}
    \begin{minipage}{\colw}\centering\scriptsize Ex4DGS\end{minipage}
    \begin{minipage}{\colw}\centering\scriptsize FTGS\end{minipage}
    \begin{minipage}{\colw}\centering\scriptsize \textbf{Ours (GraphiGS)}\end{minipage}
    \begin{minipage}{\colw}\centering\scriptsize \textbf{Ours (GraphiTS)}\end{minipage}
    \begin{minipage}{\colw}\centering\scriptsize GT\end{minipage}

    \par\vspace{4pt}
    \hrule
    \par\vspace{6pt}

    \figrow{Standard}{figs/full_page_citycrowd/citycrowd_std}
    \figrow{Sparse 50}{figs/full_page_citycrowd/citycrowd_sparse50}

    \caption{
        Qualitative comparison on our CityCrowd dataset. Each row corresponds to a different experimental setting. Columns show six competing methods and the ground truth (GT). Per-region PSNR values are reported in the top-right corner of each cropped region.
    }
    \label{fig:qualitative_page_4}
\end{figure*}

